# Exhausting Error-Prone Patterns in LDPC Codes

Chih-Chun Wang, *Member, IEEE*, Sanjeev R. Kulkarni, *Fellow, IEEE*, and H. Vincent Poor, *Fellow, IEEE*

*Abstract*— It is proved in this work that exhaustively determining bad patterns in arbitrary, finite low-density parity-check (LDPC) codes, including stopping sets for binary erasure channels (BECs) and trapping sets (also known as near-codewords) for general memoryless symmetric channels, is an NP-complete problem, and efficient algorithms are provided for codes of practical short lengths $n \approx 500$. By exploiting the sparse connectivity of LDPC codes, the stopping sets of size $\leq 13$ and the trapping sets of size $\leq 11$ can be efficiently exhaustively determined for the first time, and the resulting exhaustive list is of great importance for code analysis and finite code optimization. The featured tree-based narrowing search distinguishes this algorithm from existing ones for which inexhaustive methods are employed. One important byproduct is a pair of upper bounds on the bit-error rate (BER) & frame-error rate (FER) iterative decoding performance of arbitrary codes over BECs that can be evaluated for any value of the erasure probability, including both the waterfall and the error floor regions. The tightness of these upper bounds and the exhaustion capability of the proposed algorithm are proved when combining an optimal leaf-finding module with the tree-based search. These upper bounds also provide a worst-case-performance guarantee which is crucial to optimizing LDPC codes for extremely low error rate applications, e.g., optical/satellite communications. Extensive numerical experiments are conducted that include both randomly and algebraically constructed LDPC codes, the results of which demonstrate the superior efficiency of the exhaustion algorithm and its significant value for finite length code optimization.

## I. INTRODUCTION

Assuming iterative decoding [1], the error floor performance of arbitrary, fixed, finite low-density parity-check (LDPC) codes [2], [3] is dominated by the bad sub-structures residing in the code of interest, which are termed stopping sets when binary erasure channels (BECs) are considered [4]. For transmitting over BECs, the stopping sets not only determine the error floor performance but also determine the entire bit-error rate (BER) and the frame-error rate (FER) curves of any LDPC code [4], [5]. For general memoryless binary-input/symmetric-output channels, the dominating error-prone patterns have different names motivated by different analytical techniques; these include trapping sets [6], near-codewords [7], pseudo-codewords [8], and instantons [9]. In this work, we consider the problem of exhaustively finding the stopping sets and trapping sets (near-codewords) of any fixed finite binary linear code.

This research was supported in part by the Army Research Laboratory Collaborative Technology Alliance under Contract No. DAAD 19-01-2-0011.
This paper was presented at the 2006 IEEE International Symposium on Information Theory, Seattle, WA, USA, July. 9–14, 2006.
C.-C. Wang is with the School of Electrical and Computer Engineering, Purdue University, West Lafayette, IN 47907, USA (e-mail: chihw@purdue.edu).
S. R. Kulkarni and H. V. Poor are with the Department of Electrical Engineering, Princeton University, Princeton, NJ 08544, USA (e-mail: {kulkarni, poor}@princeton.edu).

Due to the prohibitive cost of computing the entire stopping set distribution [10] and the even greater expense of exhausting all trapping sets, in practice, the waterfall threshold of the BER is generally approximated using the density evolution [11], [12] with the corresponding scaling laws [13] for codes of different finite lengths, and pinpointed by Monte-Carlo simulation. On the other hand, the error floor can be lower bounded by identifying most of the dominant stopping sets [14], [6], by importance sampling [15], or by the instanton analysis [9], [16]. For the simplest non-trivial setting of considering the *random ensemble* of finite codes over BECs, many insightful results have been obtained including the averaged performance [4], the corresponding waterfall scaling law [13], and the ensemble error floor analysis [17], [18]. Other research related to the stopping set analysis includes, for example, results on the asymptotic stopping set weight spectrum [19], the Hamming distance spectrum [20], [21], and the stopping redundancy [22].

The minimal stopping distance is defined as the minimal size of all non-empty stopping sets. As the minimal Hamming distance is to the maximum *a posteriori* probability (MAP) detector, the minimal stopping distance is the determining factor of code performance of iterative decoders over BECs in the low error rate regimes. Recently, determining the minimal stopping distance for an arbitrary binary linear code has been proven to be an NP-complete problem [23] which shows that there is little chance that a deterministic algorithm can be devised with polynomial computational complexity Poly($n$) with respect to the codeword length $n$. Although not explicitly stated in [23], it follows that the task of determining the minimal stopping distance for LDPC codes with *sparse* parity matrices is still NP-complete regardless of the sparsity constraint on the codes of interest. By noting that for large LDPC codes, most small stopping sets are also valid codewords due to the combinatorial bias of random construction [18], this NP-completeness result is not totally unexpected since deciding the minimal Hamming distance of any linear code is one of the classic NP-complete problems of information theory [24], [25], [26]. For codes having special structure such as the Hamming codes, the LDPC codes based on circulant matrices, the convolutional LDPC codes etc., their minimal stopping / Hamming distances can be analytically determined through combinatorial or algebraic analyses [27], [28], [29], [30]. For general randomly constructed LDPC codes, no such result exists. Since the minimal stopping (or Hamming) distance is defined as the minimal size of all non-empty stopping sets (or codewords), upper bounds on the minimal stopping (or Hamming) distance can be obtained by explicitly identifying any single stopping set (or codeword) of small size, which is generally achieved by the error impulse methods [31], [32].

For non-erasure channels, the minimal trapping distance can be defined analogously as the minimal size of all non-empty trapping sets. In Section III-B, it is proved that the problem of deciding the minimal trapping distance is also NP-complete. In other words, determining the asymptotic performance of the iterative detector is not an easier task when compared to determining the asymptotic performance of the optimal maximum *a posteriori* probability detector. The NP-hardness of this problem is also one of the reasons that all existing finite-code error-floor optimization schemes employ approximations or guided heuristics as the objective functions, such as the girth of the Tanner graph [33], [34], the Approximate Cycle Extrinsic (ACE) message degree [35], partial stopping set elimination [36], and ensemble-inspired upper bounds [37]. With already significant success based on the indirect metrics, one would expect greater improvement if the minimal stopping / trapping distance can be directly employed. One purpose of this paper is to serve as a starting foundation for future research on efficient exhaustion algorithms for error-prone patterns, which can be used to decide the minimal stopping / trapping distance for corresponding code optimization.

One major motivation for this work springs from the fact that the inherent NP-hardness of the minimal stopping / trapping distance problems only describes the asymptotic complexity with large codeword length $n$, and does not preclude efficient algorithms for codes of small lengths, say $n \approx 500$, which is of great importance in practical applications. As the main contribution of this work, an efficient algorithm for exhaustively finding minimal stopping sets will be presented. By taking advantage of the sparse structure of the corresponding decoding tree, we are able to exhaust all stopping sets of size $\leq 13$ for LDPC codes of $n = 512$ and thus identify the minimal stopping distances that are $\leq 13$. For comparison, $\binom{512}{13} \approx 2.5 \times 10^{25}$ trials are required if a brute force approach is employed, which demonstrates the efficiency of our scheme especially when there is no existing exhaustion algorithm for stopping sets of non-trivial sizes other than the brute-force search. Our algorithm will be generalized for trapping sets, and we will see that all trapping sets of size $\leq 11$ can be exhausted for $n = 512$ codes. The trapping set exhaustion algorithm also explains why good codes for BECs are generally good for other channels as well.

In addition to its apparent computation-theoretic interest, the exhaustive search algorithm efficiently computes the exact value or a lower bound of the minimal stopping / trapping distance of arbitrary LDPC codes of practical sizes, which has the following three major applications:

1) For binary erasure channels, an exhaustive list of minimal stopping sets can be used to obtain a BER / FER lower bound that is tight in both *order* and *multiplicity*. For non-erasure channels, the list of small trapping sets can be used to derive lower bounds tightly predicting the error floor region [6].
2) One byproduct of our stopping set exhaustion algorithm is a BER/FER upper bound for any fixed arbitrary codes over BECs. With a pivoting rule matching the tree-based narrowing enumeration, an asymptotically tight *upper bound* can be obtained, which provides a worst performance guarantee for the regime with arbitrarily low erasure probability.
3) Using the exhausting algorithm as an objective function in finite code optimization, good codes can be obtained and the performance is guaranteed without any combinatorial outlier, a weakness of randomized constructions first pointed out in [6]. Further implementation of the code optimization algorithm and the *suppressing effects* of code lifting will be deferred to a companion paper [18].

Recent research related to the determination of the minimal Hamming distance, the pseudo-codeword weight, and the girth of the Tanner graph can be found in [38], [39], [40], [41], [42], [43]. Ensemble-based FER bounds for MAP decoders are explored in [44], [45], and the references therein. It is worth noting that although being highly correlated with the asymptotic performance [34], deciding the girth of the Tanner graph is *not* an NP-hard problem and can be easily determined with complexity $\mathcal{O}(n^2)$.

The remainder of this paper is organized as follows. This paper contains two major applications, stopping set exhaustion and trapping set exhaustion. Basic definitions of these types of error-prone patterns are given in Section II, and the NP-completeness of exhausting small trapping sets is proved in Section III. We then start from deriving a tree-based algorithm for computing the BER upper bounds for BECs in Section IV, which is then proved to be equivalent to a stopping set exhaustion algorithm based on narrowing search over the corresponding tree structure. Inspired by different properties of stopping sets and by the corresponding Boolean expression framework, several methods to further improve the efficiency of the proposed algorithm are discussed in Section V, which are the key elements of an efficient implementation. The same algorithm is generalized for trapping set exhaustion in Section VI. Section VII contains numerical experiments, including results for the (23,12,7) Golay code, the (155,64,20) Tanner code [29], the Ramanujan-Margulis ($q = 13, p = 5$) code [46], the Margulis code ($p = 7$) [47], and the two rate 1/2, $n \approx 500$ lifted LDPC codes from the companion paper [18]. Section VIII concludes this paper.

## II. PRELIMINARY & FORMULATION

### A. Code Representation

For fixed $n \in \mathbb{N}$, let $\mathbf{x} = (x_1, \ldots, x_n) \in \{0,1\}^n$ denote a transmitted vector, for which $n$ is the codeword length. The receiving signal vector denoted by $\mathbf{y} = (y_1, \ldots, y_n)$ is generally the image of the transmitted $\mathbf{x}$ corrupted by noise; $\mathsf{P}(\mathbf{y}|\mathbf{x})$ the conditional distribution of $\mathbf{y}$ given $\mathbf{x}$ describes the underlying channel. Throughout this paper, we consider only memoryless channels, i.e., $\mathsf{P}(\mathbf{y}|\mathbf{x}) = \prod_{i=1}^{n} \mathsf{P}(y_i|x_i)$. One particular type of memoryless channel is the binary erasure channel for which the received signal $y$ is either the uncorrupted transmitted signal $x$ or an erasure $e$, and a memoryless BEC can be completely described by its erasure probability $\epsilon = \mathsf{P}(y = e|x)$.

A linear code is a collection of allowed transmission vectors $C = \{\mathbf{x}\} \subseteq \{0,1\}^n$ that forms a linear subspace with addition and multiplication operations defined over the Galois field $GF(2)^n$. Any linear code $C$ can be represented by an $m \times n$ parity-check matrix $\mathbf{H}$ which contains the basis vectors of the null space of $C$, and $m$ is the dimension of the null space. I.e., if we let $\mathbf{x}$ be a column vector, then $\mathbf{Hx} = \mathbf{0}, \forall \mathbf{x} \in C$. It is worth emphasizing that this representation based on $\mathbf{H}$ is not unique. With all our results being $\mathbf{H}$-dependent, we sometimes use the code $C$ to refer to the corresponding parity-check matrix $\mathbf{H}$ instead of the collection of valid codewords $\{\mathbf{x}\}$.

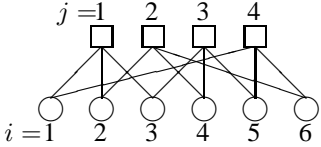

Fig. 1. A simple parity check code $C_1$.

An equivalent graphical representation of the code $C$ can be obtained by regarding the $m \times n$ matrix $\mathbf{H}$ as the incidence matrix of a bipartite graph containing two sets of vertices: the variable nodes $\{v_1, \ldots, v_n\}$ and the check nodes $\{c_1, \ldots, c_m\}$, such that $H_{ji} = 1$ if and only if there is an undirected edge connecting $(c_j, v_i)$. For example, a simple parity-check code $C_1$ with

$$\mathbf{H} = \begin{pmatrix} 1 & 1 & 1 & 0 & 0 & 0 \\ 0 & 1 & 0 & 1 & 0 & 1 \\ 0 & 0 & 1 & 1 & 1 & 0 \\ 1 & 0 & 0 & 0 & 1 & 1 \end{pmatrix}$$

can be mapped bijectively to a bipartite graph as in Fig. 1, which is termed the Tanner graph of the corresponding parity-check code. Since the coded bits $\{x_1, \ldots, x_n\}$ can be mapped to the variable nodes $\{v_1, \ldots, v_n\}$ bijectively, they are generally used interchangeably. However, in the most stringent definitions, $\{x_1, \ldots, x_n\}$ refers to the "bits" of the code vector $\mathbf{x}$ satisfying $\mathbf{Hx} = \mathbf{0}$ while $\{v_1, \ldots, v_n\}$ refers to the "nodes" in the corresponding Tanner graph.

LDPC check codes are a special family of parity check codes introduced in the early 1960s [2], [3], for which the number of ones in $\mathbf{H}$ is limited to be of the order of $n$. When combined with the message-exchanging belief propagation algorithm [48], capacity-approaching error-correction capability has been reported. Although our results are derived with the application of LDPC codes in mind, the same results hold for general parity-check codes as well.

### B. Codeword Supports, Stopping Sets & Trapping Sets

*1) Codeword Supports:* A codeword support is a subset of $\{x_1, \ldots, x_n\}$ such that if we set the values of all $x_i$'s in the codeword support to 1 and all other $x_i$'s to 0, the resulting vector $\mathbf{x}$ is a valid codeword. The Hamming weight of a codeword can thus be defined as the number of ones in the codeword or as the number of elements in the corresponding codeword support. The minimal Hamming distance $d_H$ of a parity-check code $C$ is defined as the minimal Hamming weight of all non-empty codeword supports. Since $\mathbf{Hx} = \mathbf{0}$ for all valid codewords, one can prove that a subset of $\{x_1, \ldots, x_n\}$ is a codeword support if and only if the induced subgraph contains no check nodes with odd degrees. For example, $\{x_1, x_2, x_6\}$ is a codeword support of the code $C_1$ depicted in Fig. 1.

*2) Stopping Sets:* A stopping set is a subset of $\{v_1, \ldots, v_n\}$ such that the induced subgraph contains no check nodes of degree one. Using code $C_1$ in Fig. 1 as an example, both $\{v_2, v_3, v_4\}$ and $\{v_1, v_2, v_4, v_5\}$ are stopping sets for code $C_1$. The minimal stopping distance $d_S$ is defined as the minimal size of all non-empty stopping sets. By definition, a codeword support is always a stopping set but not vice versa, which implies that $d_S \leq d_H$ for all parity-check codes. By adding redundant parity check rows to the parity-check matrix $\mathbf{H}$, one can improve the code structure so that all stopping sets are valid codeword supports and the minimal stopping distance equals the minimal Hamming distance [22].

Di *et al.* in [4] showed that for BECs, a belief propagation iterative decoder [48] fails if and only if the set of erased bits contains a stopping set. As a result, for small erasure probability $\epsilon$, the FER performance is determined by the minimal stopping distance $d_S$ and the corresponding multiplicity $m_S$, the latter of which is defined as the number of distinct stoping sets of size $d_S$.

*3) Trapping Sets:* The notion of trapping sets stems from an operational definition in which a trapping set originally refers to a subset of $\{v_1, \ldots, v_n\}$ that is susceptible to errors under message-exchanging iterative detectors in the following way. If the observations on these nodes are misleading, it is "highly unlikely" for the extrinsic messages to correct any of the nodes. The errors are "trapped" within these nodes [6]. It is a concept depending both on the underlying channel model and the decoding algorithm. For example, a stopping set is legitimately also a trapping set when the underlying channel model is the BEC and the belief propagation decoder is employed. Empirically, almost all trapping sets are *near-codewords* when channels other than BECs are considered [7]. To be more explicit, almost all trapping set are a subset of $\{v_1, \ldots, v_n\}$ such that the induced subgraph contains only a "limited" number of check nodes of odd degree and are thus "nearly" a codeword support. A near-codeword can be categorized by two numbers, the total number of variable nodes involved and the number of check nodes of odd degree in the induced graph. For example, a (12,4) near-codeword refers to a set of 12 variable nodes for which there are 4 check nodes of odd degree in the induced subgraph. To better facilitate our discussion, we redefine the trapping set from the graph-theoretic perspective, which is independent of the underlying channel model and the decoding algorithm:

*Definition 1 (k-out Trapping Sets):* A subset of $\{v_1, \ldots, v_n\}$ is a $k$-out trapping set if in the induced subgraph, there are exactly $k$ check nodes of degree one.

By the above definition, every stopping set is a 0-out trapping set. $\{v_1, v_2, v_3, v_4\}$ in $C_1$ of Fig. 1 is a 1-out trapping set. It is worth noting that the definition of a $k$-out trapping set is slightly different from the definition of near-codewords. The near-codeword focuses on the total number of check nodes of odd degree, including those with degree 1 and those with degrees $\geq 3$, and emphasizes its link to valid codewords. The $k$-out trapping set considers a pattern resembling the stopping sets but has $k$ edges connecting to check nodes of degree 1 that are able to receive correct extrinsic messages to recover the contaminated bit values. In addition to the $k$ check nodes of degree 1, there might be more check nodes of odd degrees $\geq 3$ within a $k$-out trapping set, which makes the definition different than a $(s,k)$ near codeword. This slight relaxation does not affect the generality of our exhaustion algorithm for the following two reasons. For any set of $s$ variable nodes, if it is an $(s,t)$ near-codeword, then, by definition, it must be a $k$-out trapping set for some $k \leq t$. As a result, any attempt to exhaust all $k$-out trapping sets of sizes $s \leq s_0$ and $k \leq t_0$ can also serve the purpose of exhausting all $(s,t)$ near codewords for $s \leq s_0$ and $t \leq t_0$. Furthermore, for a randomly constructed code of large length $n$, almost all small stopping sets are codewords and almost all small $k$-out trapping sets of size $s$ are $(s,k)$ near-codewords due to some combinatorial bias [18]. In our numerical examples, all minimal $k$-out trapping sets are $(s,k)$ near-codewords.

One can similarly define the minimal $k$-out trapping distance $d_{T,k}$ as the minimal size of all non-empty $k$-out trapping sets.

## III. NP-Completeness

### A. Existing Results

One important class of problems in the theory of computation is the P-class problems such that *all* problems of this type can be solved by a modern computer within (P)olynomial time Poly$(n)$ steps with respect to the input size $n$, assuming sufficient amount of memory is available. On the other hand, the class NP problem refers to those problems that are guaranteed to be solvable by a fictional (N)on-deterministic Turing machine within (P)olynomial time steps. Or equivalently, the class NP problems refer to those problems with solutions that can be "verified" by a modern computer within polynomial time steps. A class of problems is NP-hard if all other NP problems of size $n$ can be "reduced" within Poly$(n)$ time so that the algorithm of the original problem can be used as a solver for these other NP problems. There is no known efficient Poly$(n)$ time algorithm for solving *any* NP-hard problems while the existence / nonexistence of such algorithms has been one of the most important open problems within the mathematics and computer science community [49] for more than three decades. A problem is considered to be NP-complete if it is both NP and NP-hard. More rigorous definitions / discussions of these concepts can be found in [50].

Some existing NP-complete problems in the coding and information theory literature are as follows.

1) *Minimum Hamming Distance:* A binary decision problem is a special type of problem in which a yes/no question is proposed and the algorithm/solver outputs 1 if the answer to the proposed question is positive and outputs 0 if the answer is negative. We consider the following decision problem: given as input an arbitrary parity-check matrix $\mathbf{H}$ and an integer $t$, the decision problem MIN-DISTANCE$(\mathbf{H},t)$ proposes the question of whether $d_H \leq t$, in which $d_H$ is the minimal Hamming distance of the parity-check code $C$ specified by $\mathbf{H}$. It has been shown in [24], [25], [26] that the decision problem MIN-DISTANCE$(\mathbf{H},t)$ is NP-complete and the question of whether $d_H \leq t$ is hard to answer for $\mathbf{H}$ of large dimension.

Note: One can easily show that this decision problem MIN-DISTANCE$(\mathbf{H},t)$ is of similar complexity to the problem of directly computing $d_H$ given $\mathbf{H}$,[1] while the simpler nature of the Yes/No question facilitates the complexity analysis.

2) *Minimum Stopping Distance:* Given as input an arbitrary parity-check matrix $\mathbf{H}$ and an integer $t$, the decision problem STOPPING-DISTANCE$(\mathbf{H},t)$ proposes the question of whether $d_S \leq t$, in which $d_S$ is the minimal stopping distance of the parity-check code $C$ specified by $\mathbf{H}$. For notational simplicity, we use SD$(\mathbf{H},t)$ as shorthand of the decision problem STOPPING-DISTANCE$(\mathbf{H},t)$. It has been shown in [23] that SD$(\mathbf{H},t)$ is also an NP-complete problem.

One may limit the input parity-check matrix $\mathbf{H}$ to be a sparse matrix $\mathbf{H}_{\text{sparse}}$ such that the number of ones in $\mathbf{H}_{\text{sparse}}$ is of the order $\mathcal{O}(n)$. The new decision problem SD$(\mathbf{H}_{\text{sparse}},t)$ may have different complexity than the original SD$(\mathbf{H},t)$ due to this additional constraint on $\mathbf{H}$. [23] shows that SD$(\mathbf{H}_{\text{sparse}},t)$ is still an NP-complete problem regardless of such a sparsity constraint.

### B. Minimum $k$-out Trapping Distance

Given as input an arbitrary parity-check matrix $\mathbf{H}$ and an integer $t$, the decision problem $k$-OUT-TRAPPING-DISTANCE$(\mathbf{H},t)$ proposes the question of whether $d_{T,k} \leq t$, in which $d_{T,k}$ is the minimal $k$-out trapping distance of the parity-check code $C$ specified by $\mathbf{H}$. For notational simplicity, we use $k$-OTD$(\mathbf{H},t)$ as shorthand of $k$-OUT-TRAPPING-DISTANCE$(\mathbf{H},t)$. We can then prove the following theorem.

*Theorem 1:* For any fixed integer $k$, the decision problem $k$-OTD$(\mathbf{H},t)$ is NP-complete. With a sparsity constraint on the input matrix $\mathbf{H}_{\text{sparse}}$, the decision problem $k$-OTD$(\mathbf{H}_{\text{sparse}},t)$ is still NP-complete.

Before proving *Theorem 1*, we first generalize the existing NP-completeness results on deciding the minimal stopping distance. Consider a new decision problem SD$(\mathbf{H},t,v_i)$ with an additional input $v_i$. Similar to SD$(\mathbf{H},t)$, SD$(\mathbf{H},t,v_i)$ determines whether there is a non-empty stopping set containing $v_i$ and of size $\leq t$. SD$(\mathbf{H},t,v_i)$ can be thus regarded as deciding the *bit-wise* minimal stopping distance of bit $x_i$. We can further generalize the above concept and consider another decision problem SD$(\mathbf{H},t,\{v_{i_1},\ldots,v_{i_k}\})$ with the additional input being any subset of $\{v_1,\ldots,v_n\}$. Analogously, SD$(\mathbf{H},t,\{v_{i_1},\ldots,v_{i_k}\})$ decides whether there is any stopping set containing $\{v_{i_1},\ldots,v_{i_k}\}$ as a subset and of size $\leq t$. We then have the following lemma.

*Lemma 1:* All of SD$(\mathbf{H},t)$, SD$(\mathbf{H},t,v_i)$, and SD$(\mathbf{H},t,\{v_{i_1},\ldots,v_{i_k}\})$ are NP-complete problems.

---

[1]There are instances in which the decision problem is strictly easier than direct computation, for example, the primality test problem vs. the factorization problem.

**Algorithm 1** Poly-time reduction of $k$-OTD$(\mathbf{H},t)$ to SD$(\mathbf{H},t,\{v_{i_1},\ldots,v_{i_k}\})$

1: **Input:** $\mathbf{H}$ and $t$.
2: $\phi \leftarrow 0$.
3: **repeat**
4:     Based on the Tanner graph, select $k$ edges connecting to $k$ distinct check nodes and denote them as $\{(v_{i_1},c_{j_1}),\ldots,(v_{i_k},c_{j_k})\}$.
5:     **if** there is no edge between $v_{i_{k'}}$ and $c_{j_{k''}}$ for any $k' \neq k''$, **then**
6:         Construct a new $\mathbf{H}'$ by removing those columns $i$ in $\mathbf{H}$ simultaneously satisfying (1) $i \neq i_a, \forall a = 1,\ldots,k$, and (2) $\exists a \in \{1,\ldots,k\}$ such that $H_{j_a i} = 1$.
7:         Construct a new $\mathbf{H}''$ by removing row $j_a, \forall a = 1,\ldots,k$ from $\mathbf{H}'$.
8:         $\phi \leftarrow \text{SD}(\mathbf{H}'',t,\{v_{i_1},\ldots,v_{i_k}\})$.
9:     **end if**
10: **until** $\phi = 1$ or all possible selections of $k$ distinct edges are exhausted.
11: **Output:** $\phi$

**Algorithm 2** Poly-time reduction of SD$(\mathbf{H},t)$ to LESS-$k$-OTD$(\mathbf{H},t)$

1: **Input:** $\mathbf{H}$ and $t$.
2: Denote the corresponding bipartite Tanner graph by $(\{v_1,\ldots,v_n\},\{c_1,\ldots,c_m\},E)$ where $E$ contains the edges of the graph.
3: Construct a $\mathbf{H}'$ by duplicating $(k+1)$ copies of the original Tanner graph, and denote the $(k+1)n$ variable nodes as $\{v_{i,k'} : \forall i = 1,\ldots,n, \forall k' = 1,\ldots,(k+1)\}$.
4: **for** $i = 1$ to $n$ **do**
5:     Consider the $(k+1)$ duplicated variable nodes $v_{i,k'}$. Add $k(k+1)/2$ new check nodes of degree 2 and $k(k+1)$ new edges so that the shortest distance between any $v_{i,k_1}$ and $v_{i,k_2}$ is 2, and the shortest path between them is unique and uses exactly one degree 2 check node. Namely, the $(k+1)$ variable nodes now form a "clique" with the duplicated nodes linked to each other by a degree 2 check nodes.
6: **end for**
7: Denote the final graph as $\mathbf{H}''$.
8: **Output:** LESS-$k$-OTD$(\mathbf{H}'',t(k+1))$

With a sparsity constraint on the input matrix $\mathbf{H}_{\text{sparse}}$, all of SD$(\mathbf{H}_{\text{sparse}},t)$, SD$(\mathbf{H}_{\text{sparse}},t,v_i)$, and SD$(\mathbf{H}_{\text{sparse}},t,\{v_{i_1},\ldots,v_{i_k}\})$ are still NP-complete.

The above result shows that it is of similar complexity to decide the minimal stopping distance regardless of whether it is from a frame-wise perspective, from a bit-wise perspective, or from a subset-wise perspective.

*Proof:* We prove this lemma by showing that these three problems can be reduced to each other. The reduction from SD$(\mathbf{H},t)$ to SD$(\mathbf{H},t,v_i)$ is straightforward and one needs only to perform SD$(\mathbf{H},t,v_i)$ $n$ times for all $i = 1,\ldots,n$. Output 1 if $\exists i$ such that SD$(\mathbf{H},t,v_i)$ returns 1, and output 0 otherwise. The converse reduction from SD$(\mathbf{H},t,v_i)$ to SD$(\mathbf{H},t)$ is as follows. Without loss of generality, we assume the input $v_i$ being $v_n$, the last variable node. Construct a $(m+n-1) \times n$ matrix $\mathbf{H}'$ from $\mathbf{H}$ by appending $(n-1)$ rows of zeros at the bottom of $\mathbf{H}$. In the extra $(n-1)$ rows, set $H'_{j,n} = H'_{j,j-m} = 1$ for $j = m+1,\ldots,m+n-1$. With this construction, one can easily show that any stopping set with respect to the matrix $\mathbf{H}'$ must be a stopping set of $\mathbf{H}$ containing $v_n$ and vice versa. The reduction from SD$(\mathbf{H},t,v_i)$ to SD$(\mathbf{H}',t)$ is thus obtained.

The reduction from SD$(\mathbf{H},t,v_i)$ to SD$(\mathbf{H},t,\{v_{i_1},\ldots,v_{i_k}\})$ can be achieved by letting the subset be $\{v_i\}$. The converse reduction from SD$(\mathbf{H},t,\{v_{i_1},\ldots,v_{i_k}\})$ to SD$(\mathbf{H},t,v_i)$ is as follows. Construct a $(m+k) \times (n+1)$ matrix $\mathbf{H}'$ by appending an all-zero column and $k$ all-zero rows at the right and the bottom of $\mathbf{H}$ respectively. Set $H'_{j,n+1} = H'_{j,i_{j-m}} = 1$ for $j = m+1,\ldots,m+k$. With this construction, $\mathbf{v}_s \subseteq \{v_1,\ldots,v_n\}$ is a stopping set of $\mathbf{H}$ with $\{v_{i_1},\ldots,v_{i_k}\} \subseteq \mathbf{v}_s$ if and only if $\mathbf{v}_s \cup \{v_{n+1}\}$ is a stopping set of $\mathbf{H}'$ containing $v_{n+1}$. The reduction from SD$(\mathbf{H},t,\{v_{i_1},\ldots,v_{i_k}\})$ to SD$(\mathbf{H}',t+1,v_{n+1})$ is thus obtained. With the above four reductions, the proof of *Lemma 1* is complete. ∎

With the help of *Lemma 1*, the proof of *Theorem 1* can be described as follows.

*Proof of Theorem 1:* Consider an auxiliary decision problem LESS-$k$-OTD$(\mathbf{H},t)$ that decides whether there is any minimal $k_0$-out trapping distance $d_{\text{T},k_0}$ with $d_{\text{T},k_0} \leq t$ and $k_0 < k$. We first show that (i) LESS-$k$-OTD$(\mathbf{H},t)$ can be reduced to $k$-OTD$(\mathbf{H},t)$ in polynomial time, (ii) $k$-OTD$(\mathbf{H},t)$ can be reduced to SD$(\mathbf{H},t,\{v_{i_1},\ldots,v_{i_k}\})$ in polynomial time, and (iii) SD$(\mathbf{H},t)$ can be reduced to LESS-$k$-OTD$(\mathbf{H},t)$ in polynomial time.

- The first reduction (i): For any fixed $k$, the answer to LESS-$k$-OTD$(\mathbf{H},t)$ can be obtained by solving $k'$-OTD$(\mathbf{H},t)$ for all $k' < k$. Output 1 if any $k'$-OTD$(\mathbf{H},t)$, $k' < k$, returns positive, and output 0 otherwise. The correctness of this reduction is self-explanatory.
- The second reduction (ii) is described in Algorithm 1. It is straightforward to show that this reduction takes only polynomial time. Its correctness can be proved by noting that any stopping set $\mathbf{v}_s$ of $\mathbf{H}''$ containing $\{v_{i_1},\ldots,v_{i_k}\}$ must be a $k$-out trapping set of $\mathbf{H}$, and vice versa.
- The third reduction (iii) is described in Algorithm 2. The polynomial time complexity of this reduction is straightforward. The correctness of this reduction is proved as follows.

Suppose there exists a stopping set $\mathbf{v}_s$ of $\mathbf{H}$ with size $\leq t$. Let $\mathbf{v}_{s,1},\ldots,\mathbf{v}_{s,k+1}$ denote the duplicated $(k+1)$ copies of $\mathbf{v}_s$ in $\mathbf{H}''$. Then $\bigcup_{i=1}^{k+1} \mathbf{v}_{s,i}$ is a stopping set of $\mathbf{H}''$ with size $\leq t(k+1)$. LESS-$k$-OTD$(\mathbf{H}'',t(k+1))$ must return 1.

Conversely, suppose LESS-$k$-OTD$(\mathbf{H}'',t(k+1))$ returns 1. There exists a $k_0$-out trapping set $\mathbf{v}''_s$ of $\mathbf{H}''$ with size $\leq t(k+1)$ and $k_0 < k$. Without loss of generality, we may assume $v_{1,1} \in \mathbf{v}''_s$. We then focus on the clique formed by $\{v_{1,k'} : k' = 1,\ldots,k+1\}$ and the corresponding $k(k+1)/2$ degree 2 check nodes. We would like to show that all $v_{1,k'}$ must be in $\mathbf{v}''_s$. Suppose there exists a $k'$ such that $v_{1,k'} \notin \mathbf{v}''_s$. By noting that the maximal flow from $v_{1,1}$ to $v_{1,k'}$ is $k$ and by the max-flow min-cut theorem, all cuts must include at least $k$ edges. Therefore there exist $\geq k$ check nodes of degree 2 with one neighbor in $\mathbf{v}''_s$ and the other not in $\mathbf{v}''_s$. The sub-graph induced by $\mathbf{v}''_s$ must have $\geq k$ check nodes of degree 1, which leads to a contradiction. Therefore, all $v_{1,k'}$ are in $\mathbf{v}_s''$, and $\mathbf{v}_s''$ must be a union of all duplicated copies of variable nodes $\mathbf{v}_{s,i}$, namely, $\mathbf{v}_s'' = \bigcup_{i=1}^{k+1} \mathbf{v}_{s,i}$. If $\mathbf{v}_{s,i}$ is not a stopping set, then there exists at least one degree 1 check node in the subgraph of $\mathbf{H}$ induced by $\mathbf{v}_{s,i}$. Therefore, there must be at least $k+1$ degree 1 check nodes in the subgraph of $\mathbf{H}''$ induced by $\mathbf{v}_s''$, which contradicts the assumption that $\mathbf{v}_s''$ is a $k_0$-out trapping set with $k_0 < k$. From the above reasoning, $\mathbf{v}_{s,i}$ must be a stopping set of $\mathbf{H}$. The original problem $\text{SD}(\mathbf{H}, t)$ is reduced to LESS-$k$-OTD$(\mathbf{H}'', t(k+1))$.

By invoking the NP-completeness results in *Lemma 1*, *Theorem 1* is proved.[2]

∎

## IV. EXHAUSTING THE STOPPING SETS

In this section, we focus mainly on exhausting the stopping sets for the BEC while the generalization to exhausting $k$-out trapping sets will be deferred to Section VI-D.

The fact that the iterative decoder fails when the erased bits contain a stopping set implies that each stopping set $\mathbf{v}_s$ corresponds to a lower bound $\epsilon^{|\mathbf{v}_s|}$ on the FER where $\epsilon$ is the erasure probability. Exhausting stopping sets of sizes smaller than or equal to $K$ is thus equivalent to eliminating the possible existence of lower bounds of order $\mathcal{O}\left(\epsilon^{|K|}\right)$. On the other hand, an alternative way of eliminating possible FER lower bounds of order $\mathcal{O}\left(\epsilon^{|K|}\right)$ is to construct an upper bound $\mathcal{O}\left(\epsilon^{|K+1|}\right)$ if such an upper bound exists. Sharing the same effects as eliminating possible lower bounds, it can be thus shown that any FER upper bound is able to serve as an exhaustion algorithm for small stopping sets and vice versa. For the following, we construct an efficient algorithm for computing upper bounds on the error probability for arbitrary parity-check codes over BECs, which is later shown to be equivalent to an efficient stopping set exhaustion algorithm in Section V-C.

### A. The Boolean Expression Framework

Without loss of generality, we assume the all-zero codeword $\mathbf{x} = \mathbf{0}$ is transmitted for notational simplicity. With this assumption and with the BEC being considered, a decoding algorithm for bit $x_i$, $i \in \{1, \ldots, n\}$ is a function outputting either 0 or $e$, the former of which represents $x_i$ being successfully recovered / decoded and the latter of which means the bit-wise decoder fails. To be more explicit, a decoding algorithm for bit $x_i$ is a function $f_i : \{0, e\}^n \mapsto \{0, e\}$ with output $\hat{x}_i = f_i(\mathbf{y})$, where $\mathbf{y} \in \{0, e\}^n$ is the received signal vector. In this paper, $f_{i,l}$, $\forall i \in \{1, \ldots, n\}$, is used to denote the iterative decoder for bit $x_i$ after $l$ iterations. And we use $f_i = \lim_{l \to \infty} f_{i,l}$ to denote the end decoding result after an infinite number of iterations, or equivalently, the end result after the iterative decoder stops improving. If we further relabel the element "$e$" by "1," $f_i$ and $f_{i,l}$ become Boolean functions, and the BER for bit $x_i$ is simply the expectation $p_i = \mathsf{E}\{f_i(\mathbf{y})\}$. Another advantage of this conversion is that the iterative message map at the variable node then becomes "$\bullet$", the binary AND operation, and the iterative message map at the parity check node becomes "+", the binary OR operation. This Boolean-expression framework was implicitly considered in [51], [10] and many other papers discussing iterative decoders on BECs, but the power of it has never been fully utilized. Later in this section, many useful implications will be derived under this framework.

[2]The language of proving reductions herein is a simplified version of the language commonly used in the theory of computation. All our statements can be made rigorous using the formal language of reduction.

Take the simple parity-check code $C_1$ in Fig. 1 as an example. Suppose we further use $f_{i \to j,l}$ to represent the message from variable node $v_i$ to check node $c_j$ during the $l$-th iteration. The iterative decoders $f_{2,l}$, $l \in \{0, 1, 2\}$, for bit $x_2$ then become

$$f_{2,0}(\mathbf{y}) = y_2$$
$$f_{2,1}(\mathbf{y}) = y_2(y_1 + y_3)(y_4 + y_6)$$
$$\begin{aligned}f_{2,2}(\mathbf{y}) &= y_2 \left(y_1(f_{5\to 4,1} + f_{6\to 4,1}) + y_3(f_{4\to 3,1} + f_{5\to 3,1})\right) \\ &\quad \cdot \left(y_4(f_{3\to 3,1} + f_{5\to 3,1}) + y_6(f_{1\to 4,1} + f_{5\to 4,1})\right) \\ &= y_2 \left(y_1(y_5 + y_6) + y_3(y_4 + y_5)\right) \\ &\quad \cdot \left(y_4(y_3 + y_5) + y_6(y_1 + y_5)\right). \end{aligned} \quad (1)$$

As the iteration proceeds, we can derive the Boolean expression for all $f_{2,l}(\mathbf{y})$, $l \in \mathbb{N}$. The final decoder of bit $x_2$ is $f_2 := \lim_{l \to \infty} f_{2,l}$, and in this example of $C_1$, one can verify that $f_2(\mathbf{y}) = f_{2,2}(\mathbf{y})$ for all $\mathbf{y} \in \{0, 1\}^n$. Although (1) admits a beautiful nested structure, the repeated appearance of many Boolean input $y_i$'s, also known as short "cycles," poses a great challenge to the evaluation of the BER $p_2 = \mathsf{E}\{f_2(\mathbf{y})\}$. One solution is to simplify (1) by expanding the nested structure into a sum-product Boolean expression [10]:

$$f_2(\mathbf{y}) = f_{2,2}(\mathbf{y}) = y_1 y_2 y_6 + y_2 y_3 y_4 + y_1 y_2 y_4 y_5 + y_2 y_3 y_5 y_6. \quad (2)$$

$\mathsf{E}\{f_2(\mathbf{y})\}$ can then be evaluated by the projection algebra in [10] or equivalently by the inclusion-exclusion principle. For this example, we have $p_2(\epsilon) = 2\epsilon^3 + 2\epsilon^4 - 5\epsilon^5 + 2\epsilon^6$, where $\epsilon = \mathsf{E}\{y_i\}, \forall i$, is the erasure probability. For comparison, with $\epsilon = 0.1$, $p_2(\epsilon) = 2.152 \times 10^{-3}$ while the BER predicted by density evolution after 2 iterations is $\approx 1.4 \times 10^{-4}$. An order-of-magnitude gap is observed between the actual performance and the prediction by density evolution due to the presence of many short cycles.

Define the "irreducible stopping set" as a stopping set containing no other non-empty stopping set as its proper subset, which differs from the definition of the minimal stopping sets (the stopping sets of minimal size). All minimal stopping sets are irreducible stopping sets but not vice versa. It can be proved that in the simplified sum-product form, each product term can be mapped bijectively to an irreducible stopping set. For example, the product term $y_1 y_2 y_4 y_5$ in (2) corresponds to an irreducible stopping set $\{v_1, v_2, v_4, v_5\}$. The product term $y_{i_1} y_{i_2} \ldots y_{i_k}$ and the irreducible stopping set $\{v_{i_1}, v_{i_2}, \ldots, v_{i_k}\}$ will thus be used interchangeably.

Instead of constructing the exact expression of $f_i(\mathbf{y})$, in which the number of product terms is exponential with respect to $n$, if only a small collection of the product terms is identified, say $y_1y_2y_6$ and $y_2y_3y_4$, then a lower bound

$$\mathsf{E}\{f_{\mathrm{LB},2}(\mathbf{y})\} = 2\epsilon^3 - \epsilon^5 \leq \mathsf{E}\{f_2(\mathbf{y})\} = p_2(\epsilon), \quad \forall \epsilon \in [0,1]$$

can be obtained where

$$f_{\mathrm{LB},2}(\mathbf{y}) = y_1y_2y_6 + y_2y_3y_4 \leq f_2(\mathbf{y}), \quad \forall \mathbf{y} \in \{0,1\}^n.$$

In our example, $\mathsf{E}\{f_{\mathrm{LB},2}(\mathbf{y})\} = 2\epsilon^3 - \epsilon^5$ is asymptotically tight and determines the error floor. If any one of the minimal stopping sets, which are $y_1y_2y_6$ and $y_2y_3y_4$ in this example, is not identified, then the resulting lower bound will be loose. The major challenge of this approach is thus to ensure that all minimal stopping sets are exhausted. Furthermore, even when all minimal stopping sets are exhausted, this lower bound is tight only in the high signal-to-noise ratio (SNR) regime. Whether the SNR of interest is high enough can be determined only by Monte-Carlo simulations and by extrapolating the waterfall region. An upper bound thus becomes essential for a deeper understanding of the code behavior.

An upper bound can be constructed by iteratively computing the sum-product form of $f_{2,l+1}(\mathbf{y})$ from that of $f_{2,l}(\mathbf{y})$. To counteract the exponential[3] growth rate of the number of product terms with respect to $l$, during each iteration, we can "relax" and "merge" some of the product terms so that the complexity is kept manageable [10]. For example, $f_{2,2}(\mathbf{y})$ in (2) can be relaxed and merged as follows.

$$\begin{aligned} f_{2,2}(\mathbf{y}) &= y_1y_2y_6 + y_2y_3y_5y_6 + y_2y_3y_4 + y_1y_2y_4y_5 \\ &\leq 1y_2y_6 + y_21y_6 + y_21y_4 + 1y_2y_41 \\ &= y_2y_6 + y_2y_4 \\ &\triangleq f_{\mathrm{UB},2}, \end{aligned}$$

so that the number of product terms is reduced from four to two. This technique is generally very loose since the relaxation step destroys much information. To generate tight results, the number of relaxation steps must be kept minimal, which yields again an exponential growth rate of the number of product terms. As a result, good/tight results have been reported only for codes of lengths $n \leq 20$.

In contrast, we construct an efficient upper bound $\mathrm{UB}_i \geq \mathsf{E}\{f_i(\mathbf{y})\}$ by preserving much of the nested structures and tight upper bounds can be obtained for $n = 300$–$500$, a significant improvement considering the NP-hardness of this problem. Furthermore, the tightness of our bound can be verified with ease, a feature absent in the relax-and-merge approach. Combined with the lower bound $\mathsf{E}\{f_{\mathrm{LB},i}(\mathbf{y})\}$, the finite code performance can be efficiently bracketed for the first time.

### B. The Iterative Decoding Functions

Some useful notation is as follows. We use $\mathbf{y}_l^k$ to represent a partial segment $(y_l, y_{l+1}, \ldots, y_k)$ from $y_l$ to $y_k$. With this notation, $\mathbf{y} = \mathbf{y}_1^{i-1} 0 \mathbf{y}_{i+1}^n$ represents a received vector with the $i$-th value $y_i$ being 0. For any $i$, we define $y_i$ as a *determining*

[3] The corresponding exponential order is generally $\mathcal{O}\left(((d_v-1)(d_c-1))^l\right)$, where $d_v$ and $d_c$ are the maximum variable and check node degrees.

*variable* of $f(\mathbf{y})$ if $\exists \mathbf{y}_1^{i-1}, \mathbf{y}_{i+1}^n$ such that $f(\mathbf{y}_1^{i-1} 0 \mathbf{y}_{i+1}^n) \neq f(\mathbf{y}_1^{i-1} 1 \mathbf{y}_{i+1}^n)$.

*Definition 2 (Iterative Decoding Functions):* A Boolean function $f(\mathbf{y})$ is an *iterative decoding* function if $f(\mathbf{y})$ can be represented as a series of combinations of $y_1, \ldots, y_n$, the binary AND "•" and the binary OR "+" operations without using the binary NOT "¬" operation.

For example: $y_1 + y_2(y_1 + y_3)$ is an iterative decoding function while the XOR function $y_1(\neg y_2) + (\neg y_1)y_2$ is not. All $f_i(\mathbf{y})$, $f_{i,l}(\mathbf{y})$, and $f_{i \to j,l}(\mathbf{y})$ discussed in the previous section are iterative decoding functions. Before introducing a simple upper bound, we first prove the following two properties of iterative decoding functions.

*Proposition 1 (Monotonicity):* Suppose $f(\mathbf{y})$ is an iterative decoding function. Then $f(\mathbf{y})$ is monotonically increasing with respect to each input variable $y_i$. That is,

$$\forall i, \forall \mathbf{y}_1^{i-1}, \forall \mathbf{y}_{i+1}^n, \quad f(\mathbf{y}_1^{i-1} 0 \mathbf{y}_{i+1}^n) \leq f(\mathbf{y}_1^{i-1} 1 \mathbf{y}_{i+1}^n).$$

*Proof:* Since each variable node corresponds to an AND and each check node corresponds to an OR, it can be shown that any iterative decoding function can be converted to a parity-check code detection problem assuming iterative decoders. With one more uncorrupted observation $y_i$, the iterative decoder can only perform better. *Proposition 1* is simply a restatement of the channel degradation argument. This property, however, differentiates the iterative decoding function $f(\mathbf{y})$ from arbitrary Boolean functions. ∎

*Proposition 2 (Positive Correlation):* The correlation between two iterative decoding functions $f(\mathbf{y})$ and $g(\mathbf{y})$ is always non-negative, i.e., $\mathsf{E}\{f(\mathbf{y})g(\mathbf{y})\} \geq \mathsf{E}\{f(\mathbf{y})\}\mathsf{E}\{g(\mathbf{y})\}$.

*Proof:* We prove this result by induction on the number of common *determining variables* of $f(\mathbf{y})$ and $g(\mathbf{y})$. When $f(\mathbf{y})$ and $g(\mathbf{y})$ share no common *determining variable*, then $f(\mathbf{y})$ and $g(\mathbf{y})$ are independent and $\mathsf{E}\{f(\mathbf{y})g(\mathbf{y})\} = \mathsf{E}\{f(\mathbf{y})\}\mathsf{E}\{g(\mathbf{y})\}$.

Suppose the same inequality holds for $f(\mathbf{y})$ and $g(\mathbf{y})$ sharing $k$ determining variables. Consider the case in which $f(\mathbf{y})$ and $g(\mathbf{y})$ share $k+1$ determining variables. Without loss of generality, let $y_1$ be one of the common determining variables. Given $y_1 = 0$ (or $y_1 = 1$), the conditional $f(\mathbf{y})$ and $g(\mathbf{y})$ share only $k$ common determining variables. We then have

$$\begin{aligned} &\mathsf{E}\{f(\mathbf{y})g(\mathbf{y})\} \\ &= \mathsf{P}(y_1=1)\mathsf{E}\{f(\mathbf{y})g(\mathbf{y})|y_1=1\} \\ &\quad + \mathsf{P}(y_1=0)\mathsf{E}\{f(\mathbf{y})g(\mathbf{y})|y_1=0\} \\ &\geq \mathsf{P}(y_1=1)\mathsf{E}\{f(\mathbf{y})|y_1=1\}\mathsf{E}\{g(\mathbf{y})|y_1=1\} \\ &\quad + \mathsf{P}(y_1=0)\mathsf{E}\{f(\mathbf{y})|y_1=0\}\mathsf{E}\{g(\mathbf{y})|y_1=0\}, \quad (3) \end{aligned}$$

in which the last inequality follows from induction. On the other hand,

$$\begin{aligned} &\mathsf{E}\{f(\mathbf{y})\}\mathsf{E}\{g(\mathbf{y})\} \\ &= (\mathsf{P}(y_1=1)\mathsf{E}\{f(\mathbf{y})|y_1=1\} + \mathsf{P}(y_1=0)\mathsf{E}\{f(\mathbf{y})|y_1=0\}) \\ &\quad \cdot (\mathsf{P}(y_1=1)\mathsf{E}\{g(\mathbf{y})|y_1=1\} + \mathsf{P}(y_1=0)\mathsf{E}\{g(\mathbf{y})|y_1=0\}). \\ &\quad (4) \end{aligned}$$

Subtracting (4) from (3), we have

$$(3) - (4) \geq \mathsf{P}(y_1 = 1)\mathsf{P}(y_1 = 0)$$
$$\cdot (\mathsf{E}\{f(\mathbf{y})|y_1 = 1\} - \mathsf{E}\{f(\mathbf{y})|y_1 = 0\})$$
$$\cdot (\mathsf{E}\{g(\mathbf{y})|y_1 = 1\} - \mathsf{E}\{g(\mathbf{y})|y_1 = 0\}).$$

By *Proposition 1*, the right-hand side of the above inequality is always non-negative. It is thus proved that $\mathsf{E}\{f(\mathbf{y})g(\mathbf{y})\} \geq \mathsf{E}\{f(\mathbf{y})\}\mathsf{E}\{g(\mathbf{y})\}$ for $f(\mathbf{y})$ and $g(\mathbf{y})$ sharing $k+1$ common determining variables. By induction, *Proposition 2* is proved. ∎

### C. A Simple Upper Bound Based on Tree-Trimming

Let $g(\mathbf{y})$ and $h(\mathbf{y})$ be two iterative decoding functions. Consider the simple variable node message map $f_v = g \cdot h$ and the simple check node message map $f_c = g + h$ respectively. The three simplest cases are discussed as below and three rules are introduced to cope with these scenarios.

*Rule 0 — Density-Evolution-Like Evaluation*

If $g(\mathbf{y})$ and $h(\mathbf{y})$ share no determining variable, then $g(\mathbf{y})$ and $h(\mathbf{y})$ are independent and

$$\mathsf{E}\{f_v(\mathbf{y})\} = \mathsf{E}\{g(\mathbf{y})h(\mathbf{y})\} = \mathsf{E}\{g(\mathbf{y})\}\mathsf{E}\{h(\mathbf{y})\}$$
$$\mathsf{E}\{f_c(\mathbf{y})\} = \mathsf{E}\{g(\mathbf{y}) + h(\mathbf{y})\}$$
$$= \mathsf{E}\{g(\mathbf{y})\} + \mathsf{E}\{h(\mathbf{y})\} - \mathsf{E}\{g(\mathbf{y})\}\mathsf{E}\{h(\mathbf{y})\}.$$

Namely, when the inputs are independent the BER of interest can be iteratively computed by the above formulas. Nonetheless, a more interesting question is for the cases in which we have dependent $g(\mathbf{y})$ and $h(\mathbf{y})$. Can we still use this density-evolution-like (DE-like) evaluation as an upper bound? The answer to this question is addressed in Rules 1 and 2 respectively.

*Rule 1 — A Simple Relaxation*

Suppose $g(\mathbf{y})$ and $h(\mathbf{y})$ share at least one determining variable, i.e., there are repeated nodes in the input arguments of $g(\mathbf{y})$ and $h(\mathbf{y})$. By the inclusion-exclusion principle and by *Proposition 2*, we have

$$\mathsf{E}\{f_c(\mathbf{y})\} = \mathsf{E}\{g(\mathbf{y}) + h(\mathbf{y})\}$$
$$= \mathsf{E}\{g(\mathbf{y})\} + \mathsf{E}\{h(\mathbf{y})\} - \mathsf{E}\{g(\mathbf{y})h(\mathbf{y})\}$$
$$\leq \mathsf{E}\{g(\mathbf{y})\} + \mathsf{E}\{h(\mathbf{y})\} - \mathsf{E}\{g(\mathbf{y})\}\mathsf{E}\{h(\mathbf{y})\}. \quad (5)$$

The above rule suggests that when the incoming messages $g(\mathbf{y})$ and $h(\mathbf{y})$ of a check node are dependent, the error probability of the outgoing message can be upper bounded by *blindly* assuming the incoming messages are independent and by invoking Rule 0, the DE-like evaluation. Furthermore, the upper bound computed by Rule 1 has the same asymptotic order as the target error probability, as can be easily seen in (5). The multiplicity of the upper bound may be different from that of the error probability. Due to the random-like interconnection within the code graph, for most cases, $g(y)$ and $h(y)$ are "nearly independent" and the multiplicity loss is not significant. The realization of Rule 1 is illustrated in Fig. 2, in which we assume that $y_1$ is the only common determining variable.

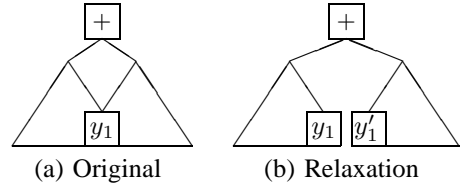

Fig. 2. Rule 1: A simple relaxation for check nodes.

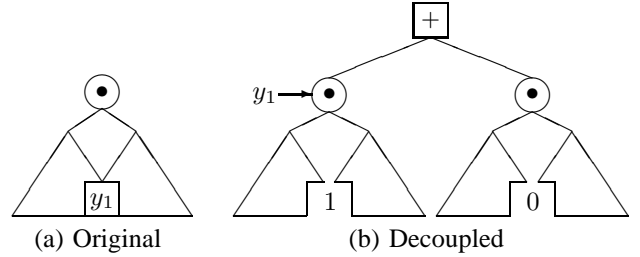

Fig. 3. Rule 2: A pivoting rule for variable nodes.

*Rule 2 — The Pivoting Rule*

Consider the simplest case in which $g(\mathbf{y})$ and $h(\mathbf{y})$ share one and only one common determining variable $y_1$. By simple Boolean algebra, we can upper bound $f_v(\mathbf{y}) = g(\mathbf{y}) \cdot h(\mathbf{y})$ as follows.

$$g(\mathbf{y}) \cdot h(\mathbf{y}) \leq g(0\mathbf{y}_2^n) \cdot h(0\mathbf{y}_2^n) + y_1 g(1\mathbf{y}_2^n) \cdot h(1\mathbf{y}_2^n).$$

By *Proposition 1*, the inequality becomes equality and we have

$$f_v(\mathbf{y}) = g(0\mathbf{y}_2^n) \cdot h(0\mathbf{y}_2^n) + y_1 \cdot g(1\mathbf{y}_2^n) \cdot h(1\mathbf{y}_2^n). \quad (6)$$

The realization of the above equation is demonstrated in Fig. 3. Once the tree in Fig. 3(a) is transformed to Fig. 3(b), messages entering the two variable nodes in Fig. 3(b) become independent since for fixed values of $y_1$, the conditional functions $g(y_1 \mathbf{y}_2^n)$ and $h(y_1 \mathbf{y}_2^n)$ share no common determining variable and are independent to each other. The expectation of the variable node AND operation can be *exactly* evaluated by Rule 0. Although the two messages entering the top check node in Fig. 3(b) are still dependent, the expectation at the check node OR operation can be *upper bounded* by reapplying Rule 1. The final upper bound thus becomes

$$\mathsf{E}\{f_v(\mathbf{y})\} \leq \mathsf{E}\{f_v(0\mathbf{y}_2^n)\} + \mathsf{E}\{y_1\}\mathsf{E}\{f_v(1\mathbf{y}_2^n)\}$$
$$- \mathsf{E}\{f_v(0\mathbf{y}_2^n)\}\mathsf{E}\{y_1\}\mathsf{E}\{f_v(1\mathbf{y}_2^n)\}, \quad (7)$$

where for all $y_1 \in \{0, 1\}$,

$$\mathsf{E}\{f_v(y_1\mathbf{y}_2^n)\} = \mathsf{E}\{g(y_1\mathbf{y}_2^n)\}\mathsf{E}\{h(y_1\mathbf{y}_2^n)\}.$$

In sum, after performing the pivoting rule as described by (6) and in Fig. 3, the expectation $\mathsf{E}\{f_v(\mathbf{y})\}$ can also be upper bounded by the DE-like evaluation.

*Remark 1:* The pivoting rule (6) is an equality and does not incur any performance loss. Any potential looseness of this upper bound (7) results from the application of Rule 1. Rule 2 thus preserves the asymptotic order of $\mathsf{E}\{f_v(\mathbf{y})\}$ as does Rule 1, and only the multiplicity term might be loose.

*Remark 2:* Although Rules 0 to 2 hold for iterative decoding functions in sum-product forms as well, it is more efficient to represent an iterative decoding function in its original tree form since the latter admits efficient computation of the DE-like evaluation. For general iterative decoding functions, the DE-like evaluation can only be used as an "approximation" of the finite code performance without concrete justification. Rules 1 and 2 provide a mechanism for converting the original tree to a suitable format such that the direct DE-like evaluation gives us a rigorous upper bound on the finite code BER.

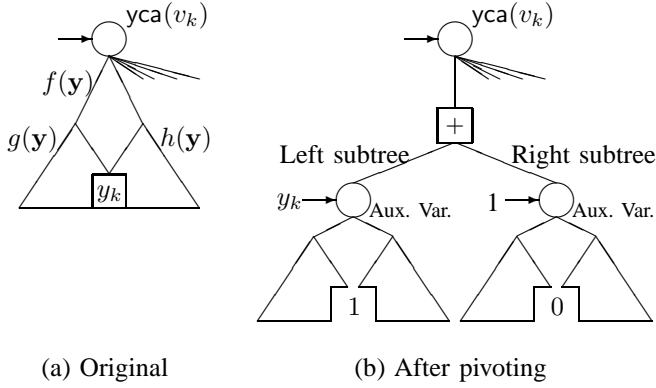

Fig. 4. The pivoting rule, Rule 2, in a real decoding tree. For the most general case, $\mathsf{yca}(v_k)$ can be an active variable node $v$ with free input $y$ or an inactive node with input hardwired to 0 or 1, which is represented by the horizontal arrow entering $\mathsf{yca}(v_k)$. $\mathsf{yca}(v_k)$ can also have $\geq 2$ children and the children not involved during pivoting are represented by lines entering $\mathsf{yca}(v_k)$ with different slopes. The function $f(\mathbf{y})$ focuses mainly on the AND operation on the two incoming messages $g(\mathbf{y})$ and $h(\mathbf{y})$ with respect to the directions of $v_k$ and $v'_k$.

### D. The Algorithm

As a simple (easy to evaluate) and powerful (tight in asymptotic order) upper bound, Rules 0 to 2 are designed to upper bound the expectation of single operations, either an AND or an OR operation, with the incoming messages $g(\mathbf{y})$ and $h(\mathbf{y})$ sharing at most one common determining variable. These apparent limitations on the size of the tree and on the number of common determining variables make them not directly applicable to real decoding problems for parity-check codes. Nonetheless, once carefully concatenated, they can be used to construct $\mathrm{UB}_i$ for the infinite-sized decoding tree with many repeated nodes while still preserving most of the nested structure. We then have the following theorem:

*Theorem 2:* Algorithm 3, a concatenation of Rules 0 to 2, is guaranteed to find a BER upper bound $\mathrm{UB}_i \geq p_i$ for the infinite-sized decoding tree of bit $x_i$.

Some technical details of this algorithm are discussed in APPENDIX I for interested readers. The proof of *Theorem 2* involves the graph theoretic properties of $\mathsf{yca}(x_j)$ and will be deferred to APPENDIX II in the interest of streamlining the discussion.

This tree-based approach corresponds to a narrowing search of stopping sets. By denoting $f_{\mathcal{T},t}(\mathbf{y})$ as the corresponding iterative decoding function based on the tree $\mathcal{T}$ after $t$ iterations of the REPEAT–UNTIL loop in Line 3, we have the following theorem.

*Theorem 3 (A Narrowing Search):* For any subset $\mathbf{x} \subseteq \{x_1, \ldots, x_n\}$, let $\mathbf{y_x} = (y_1, \ldots, y_n) \in \{0,1\}^n$ be the corresponding binary vector such that $\forall i \in \{1, \ldots, n\}$, $y_i = 1$ if

**Algorithm 3** A tree-based method for upper bounding $p_i$, the BER of bit $x_i$.

1: **Notation:** A variable node is active iff it has a *free* input observation $y$.
2: **Initialization:** Let $\mathcal{T}$ be a tree containing one active variable node $v_i$ with free input observation $y_i$ and all the neighboring check nodes of $v_i$, denoted as $c_j$. $v_i$ is used as the root of $\mathcal{T}$ and all $c_j$'s are the immediate children of $v_i$.
3: **repeat**
4:    Locate the next to-be-added variable node (also referred to as a leaf) according to the Tanner graph of the code. Suppose the next to-be-added leaf is a $v_k$ node with free observation $y_k$.
5:    Add the new leaf $v_k$ to the tree $\mathcal{T}$ and set its status to be active.
6:    Construct the immediate check node children of $v_k$ based on the Tanner graph and add them to the tree $\mathcal{T}$.
7:    **if** there exists at least another active variable node in $\mathcal{T}$ that has free input observation $y_k$ **then**
8:      Let $\mathsf{yca}(v_k)$ denote the youngest common ancestor between the newly added leaf $v_k$ and any existing active node in $\mathcal{T}$ that has free input observation $y_k$. We use $v'_k$ to denote the corresponding existing variable node in $\mathcal{T}$ such that $v_k$ and $v'_k$ jointly result $\mathsf{yca}(v_k)$.
9:      **if** $\mathsf{yca}(v_k)$ is a variable node **then**
10:        As suggested by Rule 2, a pivoting construction involving tree duplication is initiated. Use $\mathsf{yca}(v_k)$ as the root for function $f(\mathbf{y})$, the two incoming messages along the directions of $v_k$ and $v'_k$ are $g(\mathbf{y})$ and $h(\mathbf{y})$ respectively, and the shared determining variable is $y_k$. The pivoting rule with respect to $f(\mathbf{y})$, $g(\mathbf{y})$, $h(\mathbf{y})$, and $y_k$ is illustrated in Fig. 4(a–b).
11:        Deactivate all currently active variable nodes in the duplicated left and right subtrees that have free observation $y_k$ by hardwiring their free observation $y_k$ either to 1 or to 0 depending on whether the active variable node of interest is in the left subtree or in the right subtree.
12:        Use Fig. 4(b) as a reference. For the root $\mathsf{yca}(v_k)$, keep its original status. For the right duplicated subtree, deactivate its root, a newly constructed auxiliary variable node, by assigning a fixed input observation 1. For the left duplicated subtree, set the status of its root, also a newly constructed auxiliary variable node, to be active by assigning a free input observation $y_k$.
13:      **end if**
14:    **end if**
15: **until** the size of $\mathcal{T}$ exceeds the preset limit.
16: Locate all to-be-added leaf variable nodes, add them to $\mathcal{T}$, and deactivate them by hardwiring their corresponding free observation $y$ to 1.
17: $\mathrm{UB}_i$ can be computed iteratively by the formulas in Rule 0. Namely, all incoming edges are blindly assumed to be independent.

$x_i \in \mathbf{x}$ and $y_i = 0$ otherwise. Let

$$\mathbf{X}_t := \{\mathbf{x} \subseteq \{x_1, \ldots, x_n\} : f_{\mathcal{T},t}(\mathbf{y_x}) = 1\},$$

and let $\mathbf{X}_{x_i}$ denote the collection of all stopping sets containing bit $x_i$. We then have

$$\mathbf{X}_{x_i} \subseteq \mathbf{X}_{t+1} \subseteq \mathbf{X}_t, \forall t \in \mathbb{N}.$$

In other words, at each time $t$, the tree $\mathcal{T}$ represents an exhaustive collection of candidates of stopping sets containing $x_i$ and that collection narrows as iterations proceed.

The proof of *Theorem 3* is based on an incremental tree-revealing argument and is included in APPENDIX III.

We conclude this section by listing some other properties of Algorithm 3 as follows.

- The only computationally expensive step is when Rule 2 is invoked, which, in the worst case, may double the tree size and thus reduces the efficiency of Algorithm 3. Some tree-pruning rules will be introduced later in Section V-A, which alleviates the computational cost of Rule 2 significantly.

- Rule 1, being the only relaxation rule, saves much computational cost by ignoring repeated nodes. A lossless pivoting rule similar to Rule 2 can be devised based on the same principle. However, the tree structure will then be overly dissected into minute cases, which is equivalent to directly deriving the sum-product expression in (2) and is of prohibitive computational cost. The complexity-reduction of Rule 1 is a major component of the upper bounding algorithm.
- Once the tree construction is completed, evaluating $\text{UB}_i$ for any $\epsilon \in [0,1]$ can be achieved efficiently with complexity $\mathcal{O}(|\mathcal{T}|\log(|\mathcal{T}|))$, where $|\mathcal{T}|$ is the size of $\mathcal{T}$.
- The preset size limit of $\mathcal{T}$ provides a tradeoff between computational resources and the tightness of the resulting $\text{UB}_i$. Since Algorithm 3 generates a rigorous upper bound, one can terminate the program early even before the tightest results are obtained, as long as the intermediate results have met the evaluation/design requirements.

## V. Performance / Efficiency Related Issues

In this section, different schemes to further improve the efficiency/performance of $\text{UB}_i$ or to confirm the tightness of $\text{UB}_i$ will be discussed, and the end results will be compared with those of the Monte-Carlo simulations in Section VII.

### A. Pruning the Tree

To control the growth rate of the size of $\mathcal{T}$ in a manageable fashion, we introduce the following two lossless pruning rules.

*Rule 3 — The Equation:* $0 \cdot f(\mathbf{y}) = 0$

We first note that $0 \cdot f(\mathbf{y}) = 0$ for any $f(\mathbf{y})$. Therefore, if an inactive variable node $v$ is with its input observation $y$ grounded to 0, then we can completely discard all messages entering $v$. Or equivalently, we can prune all subtrees rooted at $v$ and the size of $\mathcal{T}$ will be reduced. Rule 3 can be applied to the right subtree in Fig. 4(b) every time a pivoting is performed, since the algorithm deactivates many active nodes in the right subtree by hardwiring their observations to 0. See Fig. 4 and APPENDIX I for illustration and further reference.

*Rule 4 — The Degenerate Pivoting*

We first note that the following equation holds for all iterative decoding functions $g(\mathbf{y})$.

$$y_k \cdot g(\mathbf{y}_1^{k-1} y_k \mathbf{y}_{k+1}^n) = y_k \cdot g(\mathbf{y}_1^{k-1} 1 \mathbf{y}_{k+1}^n), \quad (8)$$

for all $\mathbf{y} \in \{0,1\}^n$.

For any newly-added leaf $v_k$ with free observation $y_k$, define $\text{yca}(v_k)$ and the corresponding active variable node $v'_k$ (of free observation $y_k$) as in Algorithm 3. Motivated by the above equation, if $\text{yca}(v_k) = v'_k$, i.e., the active node $v'_k$ itself is an ancestor of the newly added leaf $v_k$, then no pivoting is necessary and we can directly deactivate $v_k$ by hardwiring its free observation $y_k$ to 1, as suggested by the right-hand side of (8).

Rule 4 can be viewed as a degenerate case of Rule 2. Suppose $\text{yca}(v_k) = v'_k$ and Rule 2 is performed, which generates the duplicated subtrees as in Fig. 4(b). Since $\text{yca}(v_k) = v'_k$, one of the two messages entering the root of the right subtree must be $y_k = 0$. By invoking Rule 3, the entire right subtree will be pruned and the original pivoting rule collapses to Rule 4.

Whether $\text{yca}(v_k) = v'_k$ should be checked every time a pivoting is about to be performed. If $\text{yca}(v_k) = v'_k$, then we can use Rule 4 instead of pivoting, and the computational cost is thus reduced.

These two rules help the algorithm to use its computational resources in a more efficient way, especially the use of memory. The pruning rules also play an important role during the proof of the optimality of Algorithm 3.

### B. The Leaf-Finding Module

The tightness of $\text{UB}_i$ in Algorithm 3 depends heavily on the leaf-finding (LF) module invoked in Line 4. From the perspective of narrowing the set of potential stopping sets, the LF module determines the reduction from $\mathbf{X}_t$ to $\mathbf{X}_{t+1}$ during each iteration. A properly designed LF module is capable of resulting in an upper bound $\text{UB}_i$ of which the asymptotic order is +1 to +3 better than that of a random LF module. To further illustrate the importance of the LF module, the ultimate benefit of an optimal LF module is stated in the following theorem.

*Theorem 4 (The Optimal LF Module):* Following the notation in *Theorem 3*, with an optimal LF module, the simple pruning rules in Section V-A, and an infinite amount of memory, there exists $t_0 < \infty$ such that

$$\mathbf{X}_{x_i} = \mathbf{X}_t, \quad \forall t \geq t_0.$$

In words, our narrowing search is able to implicitly identify *all* stopping sets containing $x_i$ through the constructed tree $\mathcal{T}$, which is of a special structure that will be used later for computing the upper bound $\text{UB}_i$. Furthermore, this implicit identification can be achieved within a finite number of steps.

*Theorem 4* also demonstrates the importance of a properly designed LF module and the pruning rules, namely, if a suboptimal LF module is used or the pruning rules are not employed, $\mathbf{X}_t$ may be strictly bounded away from the collection of all stopping sets $\mathbf{X}_{x_i}$, as $t \to \infty$, and a huge performance loss is expected. The proof of *Theorem 4* is provided in APPENDIX IV.

*Corollary 1 (Order Tightness of Algorithm 3):* When combined with an optimal LF module and the pruning rules, the $\text{UB}_i$ computed by Algorithm 3 is tight in terms of the asymptotic order if sufficient computational resources are provided. Namely, $\exists C > 0$ such that $\frac{\text{UB}_i(\epsilon)}{p_i(\epsilon)} < C$ for all $\epsilon \in (0,1]$.

  *Proof:* This is a direct result of *Theorem 4* and the order preserving property of Rule 1, the only relaxation rule. To be more explicit, *Theorem 4* implies $f_{\mathcal{T},t}(\mathbf{y}) = f_i(\mathbf{y})$ for all $t \geq t_0, \mathbf{y} \in \{0,1\}^n$. By *Proposition 4* in APPENDIX II, the proof is complete. ∎

For all our experiments, an efficient approximation of the optimal LF module, motivated by the proof of *Theorem 4*, is adopted. Running on a personal computer, Algorithm 3 is capable of constructing asymptotically tight upper bounds for rate-1/2 LDPC codes of $n \leq 100$ and of minimal stopping distances 9–11. A composite approach will be introduced later, which further extends the applicable range, and asymptotically tight upper bounds can be obtained for codes with $n \approx 300$–$500$ and minimal stopping distances 11–13.

## C. Confirming the Tightness of $\mathrm{UB}_i$: An Exhaustive List of Minimum Stopping Sets

Two important features of the proposed algorithm are the easy confirmation of the tightness of $\mathrm{UB}_i$ and a byproduct tight lower bound $\mathrm{LB}_i$, which together with $\mathrm{UB}_i$ bracket the performance of the parity-check code of interest. In this subsection, we will also discuss how to construct an exhaustive list of minimal stopping sets containing $x_i$ so that Algorithm 3 can be used as an algorithm for determining the minimal stopping distance from the bit $x_i$ perspective.

To this end, after $t$ iterations of Algorithm 3, we first exhaustively enumerate the elements of minimal weight in $\mathbf{X}_t$. Denote the weight as $w_\mathcal{T}$ and denote the collection of the minimal weight elements as $\mathbf{X}_{\min} \subseteq \mathbf{X}_t$.

*Corollary 2 (Tightness Confirmation):* If $\exists \mathbf{x} \in \mathbf{X}_{\min}$ that is a stopping set, then $\mathrm{UB}_i$ is tight in terms of the asymptotic order. Otherwise, $\mathrm{UB}_i$ is loose.

*Corollary 3 (Stopping Set Exhaustion):* Let $\mathbf{X}_{\min,\mathrm{SS}} \subseteq \mathbf{X}_{\min}$ denote the collection of all elements $\mathbf{x} \in \mathbf{X}_{\min}$ that are also stopping sets. If $\mathbf{X}_{\min,\mathrm{SS}}$ is not empty, then $\mathbf{X}_{\min,\mathrm{SS}}$ exhausts all minimal stopping sets containing bit $x_i$. If $\mathbf{X}_{\min,\mathrm{SS}} = \emptyset$, then there is no stopping set of size $\leq w_\mathcal{T}$.

*Corollary 4 (The Tight Lower Bound):* A non-empty $\mathbf{X}_{\min,\mathrm{SS}}$ can be used to derive a lower bound $\mathsf{E}\{f_{\mathrm{LB},i}(\mathbf{y})\}$ that is guaranteed to be tight in both the asymptotic order and the multiplicity as described in Section IV-A.

*Proofs of Corollaries 2 to 4:* We first notice that $w_\mathcal{T}$ is the asymptotic order of both $\mathsf{E}\{f_\mathcal{T}(\mathbf{y})\}$ and the upper bound $\mathrm{UB}_i$ due to the order-preserving property of Rule 1. Suppose $\exists \mathbf{x} \in \mathbf{X}_{\min}$ that is a stopping set. By *Theorem 3*, no stopping set containing $x_i$ has weight less than $w_\mathcal{T}$ and the asymptotic order of $p_i = \mathsf{E}\{f_i(\mathbf{y})\}$ is $w_\mathcal{T}$. $\mathrm{UB}_i$ is thus an upper bound of $\mathsf{E}\{f_i(\mathbf{y})\}$ tight in the asymptotic order. For the other case that no element of $\mathbf{X}_{\min}$ is a stopping set, by *Theorem 3*, any stopping set containing $x_i$ must have weight strictly larger than $w_\mathcal{T}$ and therefore the asymptotic order of $\mathsf{E}\{f_i(\mathbf{y})\}$ must be strictly larger than $w_\mathcal{T}$. $\mathrm{UB}_i$ is loose and *Corollary 2* is proved.

By *Theorem 3*, the collection of all stopping sets containing $x_i$ is a subset of $\mathbf{X}_t$. If $\exists \mathbf{x} \in \mathbf{X}_{\min}$ that is a stopping set, then $\mathbf{X}_{\min,\mathrm{SS}}$ is not empty and all minimal stopping sets must be elements of $\mathbf{X}_{\min}$ and $\mathbf{X}_{\min,\mathrm{SS}}$. If $\mathbf{X}_{\min,\mathrm{SS}}$ is empty, any stopping set containing $x_i$ must have weight larger than $w_\mathcal{T}$. *Corollary 3* is proved.

*Corollary 4* is then a straightforward result from the exhaustive list of minimal stopping sets containing $x_i$. ∎

The exhaustive list $\mathbf{X}_{\min,\mathrm{SS}}$ gives us a richer understanding of the code behavior than the Yes/No answer to the decision problem $\mathrm{SD}(\mathbf{H}, t, x_i)$ in Section III-B, the latter of which gives only the existence result while the former of which explicitly locates all minimal stopping sets involving $x_i$.

## VI. Further Generalizations

In the previous sections, a tree-based algorithm for upper bounding BER performance and exhausting small stopping sets is provided. To further demonstrate the versatility of the proposed algorithm, we discuss the following generalizations, including the $k$-out trapping set exhaustion algorithms.

### A. Applications on Shortened and Punctured Codes

Algorithm 3 is originally designed for arbitrary parity-check codes but can be easily generalized for codes with punctured or shortened bits as follows.

If we take a closer look at the puncture operation on a particular bit $x_k$, it is equivalent to deterministically assigning "erasure" to $y_k$. I.e., it becomes an erasure channel with $y_k = 1$ with probability one. Similarly, shortening[4] a bit $x_k$ is equivalent to deterministically assigning value 0 to the observation $y_k$. Based on this observation, Algorithm 3 can be applied to codes with punctured and shortened bits by deactivating the newly added leaf $v_k$ in Line 5 and hardwiring its corresponding $y_k$ to either 1 or 0 depending on whether $v_k$ is a punctured bit or a shortened bit. The ability of analyzing shortened / punctured codes makes the algorithm applicable to irregular/regular repeat-accumulate codes as well.

Not only can the algorithm deal with the punctured and the shortened bits, the upper bound $\mathrm{UB}_i$ can be computed for the setting of parallel independent channels, in which different bits $x_k$ may experience different erasure probabilities $\epsilon_k = \mathsf{E}\{y_k\}$. No modification of Algorithm 3 is necessary since Algorithm 3 does not assume a common erasure probability for different bits $x_k$.

### B. The Composite Approach

One brute force method of searching for small stopping sets is to exhaustively consider all possible candidates of small stopping sets and test them one by one to see whether any of them is a stopping set, the complexity of which grows exponentially for large $n$. Although completely impractical, the divide-and-conquer approach of brute-force search can be combined with our tree-based algorithm. A more efficient composite approach can thus be obtained as follows.

We first notice that the expectation $\mathsf{E}\{f_i(\mathbf{y})\}$ can be further decomposed as

$$\mathsf{E}\{f_i(\mathbf{y})\} = \sum_{j=1}^{M} \mathsf{P}(\mathcal{A}_k) \mathsf{E}\{f_i(\mathbf{y})|\mathbf{y} \in \mathcal{A}_k\},$$

where the $\mathcal{A}_k$'s are $M$ events partitioning the sample space $\{0,1\}^n$. For example, we can define a collection of non-uniform $\mathcal{A}_k$'s by

$$\begin{aligned} \mathcal{A}_1 &= \{\mathbf{y} \in \{0,1\}^n : y_{10} = 0\} \\ \mathcal{A}_2 &= \{\mathbf{y} \in \{0,1\}^n : y_{10} = 1, y_{47} = 0\} \\ \mathcal{A}_3 &= \{\mathbf{y} \in \{0,1\}^n : y_{10} = 1, y_{47} = 1\}. \end{aligned}$$

Since for any $k$, $f_i(\mathbf{y})|_{\mathbf{y} \in \mathcal{A}_k}$ is simply another finite code with many punctured and shortened bits, Algorithm 3 can be applied to each $f_i(\mathbf{y})|_{\mathbf{y} \in A_k}$ respectively and different bounds $\text{UB}_{i,k} \geq \mathsf{E}\{f_i(\mathbf{y})|\mathbf{y} \in A_k\}$ will be obtained. A composite upper bound can then be constructed by

$$\text{C-UB}_i = \sum_{k=1}^{M} \mathsf{P}(\mathcal{A}_k)\text{UB}_{i,k} \geq \mathsf{E}\{f_i(\mathbf{y})\} = p_i.$$

---

[4]The term "shortening" generally refers to hardwiring some coded bits and not sending them through the channel, the effect of which is that both the message and the codeword spaces are reduced. In this work, we use the term "shorten a bit $x_k$" to stress that bit $x_k$ is hardwired to zero and is not sent through the channel.

In the case that Algorithm 3 is used as a stopping set exhaustion algorithm, a composite exhaustion algorithm can be obtained by assuming the following slight adjustment. Consider the event $\mathcal{A}_2$ as an example. Applying Algorithm 3 to $f_i(\mathbf{y})|_{\mathbf{y} \in \mathcal{A}_2}$ returns an exhaustive list of the minimal stopping sets of the corresponding punctured/shortened code in which $x_{10}$ is punctured while $x_{47}$ is shortened. The resulting minimal stopping set $\mathbf{x}$ however will not include $x_{10}$ since $y_{10} = 1$ is considered to be a fixed, punctured structure of the code instead of a free observation. By taking the union, $\{x_{10}\} \cup \mathbf{x}$ becomes a minimal stopping set of the original code $f_i(\mathbf{y})$ with $\mathbf{y}$ being confined to $\mathcal{A}_2$:

$$f_i(\mathbf{y}_{\{x_{10}\} \cup \mathbf{x}}) = 1, \text{ and } \mathbf{y}_{\{x_{10}\} \cup \mathbf{x}} \in \mathcal{A}_2.$$

In sum, when the minimal stopping set of interest is the union of the punctured bits and the conditional minimal stopping set $\mathbf{x}$ returned by Algorithm 3, the composite approach is able to exhaustively search through the partitioned space $\bigcup_{k=1}^{M} \mathcal{A}_k$ one event at a time. The reduced sample space in each event $\mathcal{A}_k$ makes the composite approach more efficient when compared to the original algorithm.

With a properly chosen non-uniform partition $\{\mathcal{A}_k\}$ of 50–20,000 events, the composite approach generally results in upper bounds of the asymptotic order +1 to +3 higher than the original $\text{UB}_i$, and is capable of exhausting stopping sets of sizes +1 to +3 larger than the capability of the original algorithm. With the improved efficiency, our algorithm is then able to determine the minimal stopping distance and identify all minimal stopping sets for many rate-1/2 LDPC codes of practical length $n = 512$.

### C. BER vs. FER

Thus far, all our discussions have been based on the bit-wise perspective. We either construct an upper bound on the BER $p_i$, or exhaust all minimal stopping sets involving bit $x_i$. Upper bounds for the average BER can be easily obtained by taking averages over bounds for individual bits. An equally interesting problem is to upper bound the FER. In this subsection, we briefly discuss how to upper bound the FER or to exhaust the minimal stopping sets from the frame perspective.

We again rely on the Boolean expression framework. By noticing that the frame error detector is simply the binary OR of all individual bit detectors, we have $f_{\text{FER}}(\mathbf{y}) = \sum_{i=1}^{n} f_i(\mathbf{y})$, which is yet another iterative decoding function. Since Algorithm 3 is applicable to all iterative decoding functions, the upper bound $\text{UB}_{\text{FER}}$ and the exhaustive list of minimal stopping sets can be obtained by direct application of Algorithm 3 on $f_{\text{FER}}(\mathbf{y})$ instead of the individual $f_i(\mathbf{y})$. A graphical interpretation can be obtained by introducing an auxiliary variable and check node pair $(x_0, y_0)$ such that the new variable node $x_0$ is punctured and the new check node $y_0$ is connected to all $n+1$ variable nodes from $x_0$ to $x_n$. The FER of the original code now equals the BER $p_0$ of variable node $x_0$ and can be upper bounded by Algorithm 3. An efficient and straightforward partition $\{\mathcal{A}_k\}_{k \in \{1,\dots,n+1\}}$ for the FER detection $f_{\text{FER}}(\mathbf{y})$ is to let

$$\forall k \in \{1,\dots,n\}, \ \mathcal{A}_k := \{\mathbf{y} : \mathbf{y}_1^{k-1} = \mathbf{0}, y_k = 1\},$$
$$\text{and} \quad \mathcal{A}_{n+1} := \{\mathbf{y} : \mathbf{y} = \mathbf{0}\}. \quad (9)$$

All our FER results are based on the above partition or on a finer partition derived from the above basic partition.

*Remark:* Since the FER depends only on the worst bit performance, it is generally easier to construct tight $\text{UB}_{\text{FER}}$ for the FER than for the BER. One might thus question the need of bit-wise upper bounds / exhaustive lists of minimal stopping sets, which on the other hand, provide detailed performance prediction for each individual bit and is of great importance during code optimization and analysis [36], [18].

### D. Exhausting $k$-Out Trapping Sets

*The Algorithm*

We are now ready to generalize the bit-wise minimal stopping set exhaustion algorithm for $k$-out trapping sets in a frame-wise perspective. We start with converting our bit-wise stopping set exhaustion algorithm to its frame-wise counterpart. Namely, using the procedures described in the previous sections, one is able to obtain an efficient algorithm taking inputs $\mathbf{H}$, $t$, and a list of punctured bits[5] $\mathbf{x}_p = \{x_p\}$, and outputting $(\mathbf{X}_{\min}, w_{\min})$. If the minimal size of the stopping sets is within the searchable range $t$, $\mathbf{X}_{\min}$ is an exhaustive list of the frame-wise minimal stopping sets and $w_{\min}$ is the corresponding minimal size. If the stopping distance is beyond the searchable range $t$, then $\mathbf{X}_{\min} = \emptyset$ and $w_{\min} = t$. We denote this algorithm as the minimal stopping set exhaustion $\text{SSE}(\mathbf{H}, t, \mathbf{x}_p)$. The minimal $k$-out trapping set exhaustion algorithm is then described in Algorithm 4, the correctness of which follows from the reduction stated in Algorithm 1 in Section III-B.

The complexity of Algorithm 4, although being polynomial with respect to $n$, grows on the order of $\mathcal{O}(n^k)$, which makes the above algorithm less appealing for cases of large $k$. Using the efficient stopping set exhaustion algorithm $\text{SSE}(\mathbf{H}, t, \mathbf{x}_p)$ proposed in this paper, meaningful results, i.e., with the search range $t$ large enough for short practical codeword length $n \approx 500$, have been reported for the cases of $k = 1, 2$.

On the other hand, the dominating error patterns are generally those $k$-out trapping sets with small $k \leq 4$, as noted in [6]. In the same paper, it has been shown that for a rate 51/64, $n = 4096$ almost-regular code example, more than 55% of the FER is contributed by the minimal $k$-out trapping sets with $k = 0, 1, 2$, equivalent to the (10,0), (9,1), (8,2) near-codewords respectively [Figure 5, [6]]. Being capable of exhausting minimal $k$-out trapping sets with $k \leq 2$, Algorithm 4 is a successful first attempt on the inherently NP-complete

---

[5]The shortening effect can be easily incorporated by removing the corresponding columns in $\mathbf{H}$ and thus will not be considered separately.

**Algorithm 4** A minimal $k$-out trapping set exhaustion algorithm based on the stopping set exhaustion algorithm SSE($\mathbf{H}, t, \mathbf{x}_p$)

1: **Input:** $\mathbf{H}$ and $t$.
2: **Initialization:** $\mathbf{X}_{\min} \leftarrow \emptyset$ and $w_{\min} \leftarrow t$.
3: **repeat**
4:   Based on the Tanner graph, select $k$ edges connecting to $k$ distinct check nodes and denote them as $\{(x_{i_1}, c_{j_1}), \ldots, (x_{i_k}, c_{j_k})\}$.
5:   **if** there is no edge between $x_{i_{k'}}$ and $c_{j_{k''}}$ for any $k' \neq k''$, **then**
6:     Construct a new $\mathbf{H}'$ by removing those columns $i$ in $\mathbf{H}$ simultaneously satisfying (1) $i \neq i_a, \forall a = 1, \ldots, k$, and (2) $\exists a \in \{1, \ldots, k\}$ such that $H_{j_a i} = 1$.
7:     Construct a new $\mathbf{H}''$ by removing rows $j_a, \forall a = 1, \ldots, k$ from $\mathbf{H}'$.
8:     Let the punctured bits be $\mathbf{x}_p \leftarrow \{x_{i_a} : \forall a = 1, \ldots, k\}$
9:     $(\mathbf{X}_{\text{temp}}, w_{\text{temp}}) \longleftarrow$ SSE($\mathbf{H}'', t, \mathbf{x}_p$).
10:     **if** $w_{\text{temp}} + k < w_{\min}$ **then**
11:       $\mathbf{X}_{\min} \leftarrow \mathbf{X}_{\text{temp}} \circ \mathbf{x}_p$† and $w_{\min} \leftarrow w_{\text{temp}} + k$.
12:     **else if** $w_{\text{temp}} + k = w_{\min}$ **then**
13:       $\mathbf{X}_{\min} \leftarrow \mathbf{X}_{\min} \cup (\mathbf{X}_{\text{temp}} \circ \mathbf{x}_p)$.†
14:     **end if**
15:   **end if**
16: **until** all possible selections of $k$ distinct edges are exhausted.
17: **Output:** $\mathbf{X}_{\min}$ and $w_{\min}$.

†: The "∘" operation is defined as follows. Any element in $\mathbf{X} \circ \mathbf{x}_p$ is of the form $\mathbf{x} \cup \mathbf{x}_p$ for some $\mathbf{x} \in \mathbf{X}$ and vice versa.

problem of trapping set exhaustion. More numerical examples will be reported in Section VII, including the exhaustion of all minimal 2-out trapping sets of size 8 for the (155,64,20) Tanner code [29].

*Good codes for BECs are good for other channels*

It is observed in [34] that good codes (in terms of error floors) for BECs are generally good for other channels as well. As a byproduct of Algorithm 4, *Theorem 5* provides a justification of this phenomenon as follows.

*Definition 3:* A code is "$(d_S, \#c_M)$ uniformly good" for BECs if after removing any $\#c \leq \#c_M$ check nodes, the resulting code has minimal stopping distance no less than $d_S$.

*Theorem 5:* A $(d_S, \#c_M)$ uniformly good code (for BECs) has minimal $k$-out trapping distance no less than $d_S$ for all $k \leq \#c_M$.

*Proof:* This is a straightforward result of the correctness of Algorithm 4. ∎

Codes constructed based on girth optimization are generally uniformly good for BECs, since removing any check node will only increase the girth, which is the lower bound of the minimal stopping distance. *Theorem 5* guarantees that any uniformly good codes for BECs are also good for other channels since they have good $k$-out trapping distances as well.

*Some Remarks*

As the key building block of trapping set exhaustion, the minimal stopping set exhaustion SSE($\mathbf{H}, t, \mathbf{x}_p$) plays a central role in the exhaustion of error-prone patterns, including both the stopping and trapping sets, and deserves the attention of future research along this direction. Two final remarks on the efficiency and algorithmic issues are as follows.

*Remark 1:* The core of our results is the bit-wise Algorithm 3. Due to its bit-based characteristics, the efficiency of our algorithms can be improved for codes of special structure by using the corresponding automorphism, including the Margulis codes, the Ramanujan-Margulis codes, and the lifted codes based on permutation matrices, etc. Taking the cyclically lifted codes with lifting factor $L$ as an example, the complexity of Algorithm 4 for $k$-out trapping sets can be reduced by a factor of $2^{k-1}L$, which makes our algorithm especially suitable for the protograph codes that are based on small base codes with high lifting factor $L$ [52].

*Remark 2:* For large $k$, exhausting $k$-out trapping sets becomes tricky especially when the minimal variable node degree is $\leq k$. If a single variable node $v$ is of degree $\deg(v) \leq k$, then $\{v\}$ itself is the minimal $k$-out trapping set and would thus dominate the outputs of Algorithm 4 and make the results useless. For example, searching minimal 2-out trapping sets of an irregular code with degree 2 nodes always returns trapping sets containing single degree 2 nodes. A simple remedy is to carefully select the $\{(x_{i_1}, c_{j_1}), \ldots, (x_{i_k}, c_{j_k})\}$ of interest in Line 4 of Algorithm 4 and preclude the uninteresting combinations. A more effective method would require unambiguous definitions of 'non-trivial' $k$-out trapping sets and is currently under investigation.

## VII. NUMERICAL EXPERIMENTS

The experiments will be divided into four categories: (i) non-sparse codes, (ii) LDPC codes with random construction, (iii) error-floor optimized LDPC codes, and (iv) algebraically constructed codes. We use SSE and $k$-TSE as shorthand of the minimal stopping set exhaustion and the minimal $k$-out trapping exhaustion algorithms discussed in the previous sections.

### A. Non-Sparse Codes

*1) The (23,12) Binary Golay Code:* The standard parity check matrix of the Golay code is described by $\mathbf{H} = [\mathbf{H}' \; \mathbf{I}]$, where $\mathbf{I}$ is an $11 \times 11$ unity matrix and $\mathbf{H}'$ is as follows.

$$\mathbf{H} = [\mathbf{H}' \; \mathbf{I}], \quad \mathbf{H}' = \begin{pmatrix} 100111000111 \\ 101011011001 \\ 101101101010 \\ 101110110100 \\ 110011101100 \\ 110101110001 \\ 110110011010 \\ 111001010110 \\ 111010100011 \\ 111100001101 \\ 011111111111 \end{pmatrix}.$$

The minimal stopping distance of the Golay code is 4 and all 130 minimal stopping sets can be found by the proposed SSE in two seconds. All bits are involved in at least one stopping set of size 4, and thus their BERs are of similar magnitudes. Fig. 5 compares the upper bound (UB), the composite upper bound (C-UB), the Monte-Carlo simulation (MC-S), and the tight lower bound (LB), for bits 0, 5, and 20. As illustrated, C-UB and LB tightly bracket the MC-S results, which shows that the UB and C-UB are capable of decoupling even *non-sparse* Tanner graphs with plenty of cycles. However, the non-sparsity slows down the efficiency of SSE considerably when

TABLE I

| (a) $n = 50$ | | | | | |
|---|---|---|---|---|---|
| Order | 3 | 4 | 5 | 6 | 7 |
| Num. bits | 3 | 11 | 10 | 20 | 6 |
| order* | | | 3 | 8 | 5 |
| + multi* | 3 | 11 | 7 | 12 | 1 |

| (b) $n = 72$ | | | | | |
|---|---|---|---|---|---|
| Order | 2 | 4 | 5 | 6 | 7 | 8 |
| Num. bits | 4 | 4 | 5 | 28 | 28 | 3 |
| order* | | | 1 | 11 | 26 | 1 |
| + multi* | 4 | 4 | 4 | 17 | 2 | |

| (c) $n = 144$ | | | | | |
|---|---|---|---|---|---|
| Order | 2 | 5 | 7 | 8 | 9 |
| order* | 4 | 3 | 7 | 27 | 2 |
| order> | | | 6 | 90 | 5 |

PERFORMANCE STATISTICS FOR COMPLETE RANDOMLY CONSTRUCTED (3,6) LDPC CODES: "Num. bits" is the number of bits with the specified asymptotic order. "order*" is the num. bits with UBs tight only in the order. "+ multi*" is the num. bits with UBs tight both in the order and in the multiplicity. "order >" is the num. bits with a UB of the specified order while no bracketing lower bound can be established.

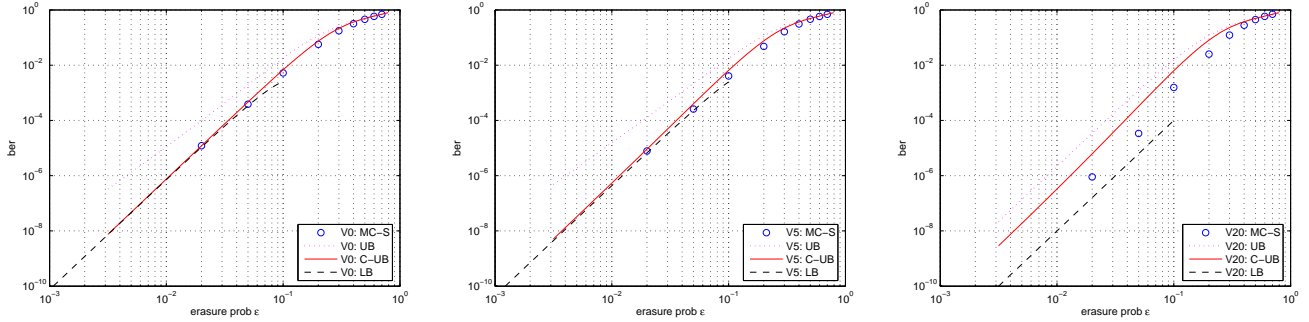

Fig. 5. Comparisons among the upper bound (UB), the composite upper bound (C-UB), the Monte-Carlo simulation (MC-S), and the side product tight lower bound (LB) for bits 0, 5, and 20 of the (23,12) binary Golay code. The corresponding asymptotic (order, multiplicity) pairs obtained by UB, C-UB, and the actual BER is are V0: $\{(3, 10), (4, 75), (4, 75)\}$, V5: $\{(3, 15), (4, 50), (4, 45)\}$, and V20: $\{(4, 221), (4, 27), (4, 1)\}$.

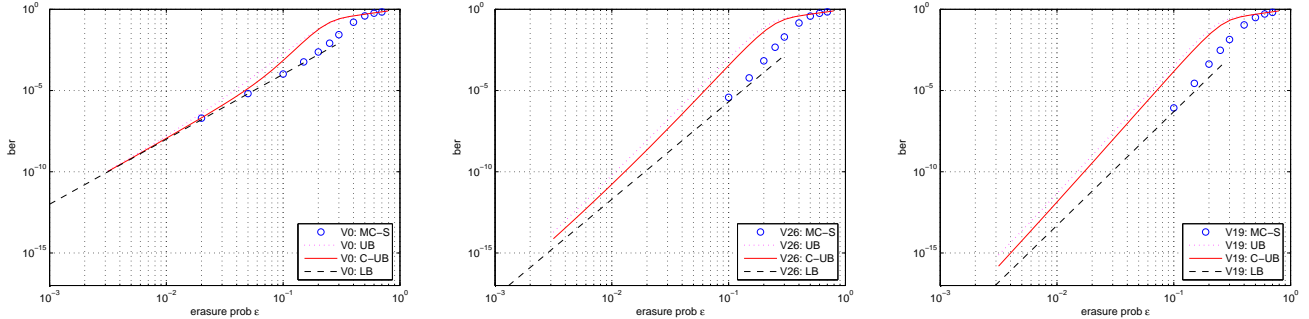

Fig. 6. Comparisons among the UB, the C-UB, the MC-S, and the LB for bits 0, 26, and 19 of a randomly generated (3,6) LDPC code with $n = 50$. The asymptotic (order, multiplicity) pairs of the UB, the C-UB, and the actual BER are V0: $\{(4, 1), (4, 1), (4, 1)\}$, V26: $\{(6, 2), (6, 4), (6, 2)\}$, and V19: $\{(7, 135), (7, 10), (7, 5)\}$.

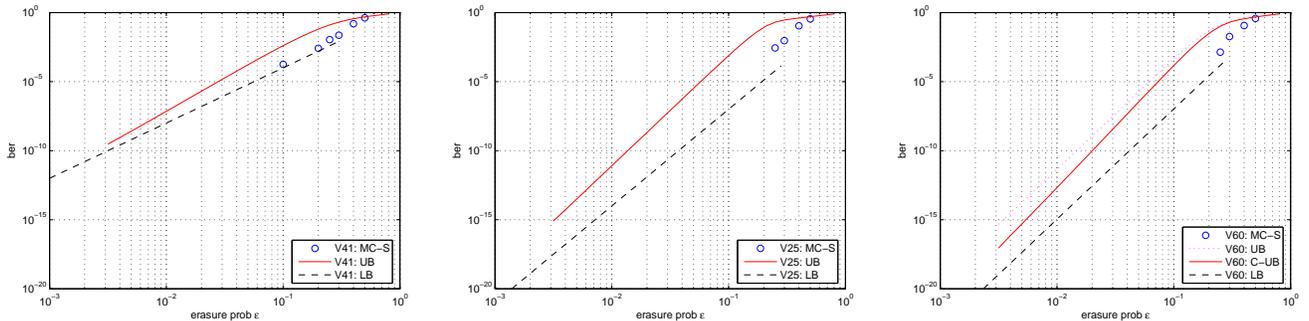

Fig. 7. Comparisons among the UB, the C-UB, the MC-S, and the LB for bits 41, 25, and 60 of a randomly generated (3,6) LDPC code with $n = 72$. The asymptotic (order, multiplicity) pairs of the UB, the C-UB, and the actual BER are V41: $\{(4, 1), -, (4, 1)\}$, V25: $\{(7, 1), -, (7, 1)\}$, and V60: $\{(7, 19), (8, 431), (8, 11)\}$.

compared to sparse LDPC codes of similar sizes, due to the higher frequency of invoking the pivoting rule. Nonetheless, the proposed SSE still demonstrates superior performance when compared to brute force search. For comparison, the extremely short codeword length of the Golay code makes the stopping set exhaustion problem still within the reach of the brute-force search, which requires $\binom{23}{4} = 490,314$ trials.

### B. LDPC Codes with Random Construction

*1) A (3,6) LDPC Code with $n = 50$:* A (3,6) LDPC code with $n = 50$ is randomly generated, and the UB, the C-UB, the MC-S, and the tight LB are performed on bits 0, 26, and 19, as plotted in Fig. 6, and the statistics of all 50 bits are provided in TABLE I(a). The UB is tight in the asymptotic order for *all* bits while for 34 bits, the UB is also tight in multiplicity. Among the 16 bits for which UB is not tight in multiplicity, 11 bits are within a factor of three of the actual multiplicity, which is obtained from the bit-wise exhaustive list of stopping sets obtained from the SSE.

In contrast with the Golay code example, the tight performance can be attributed to the sparse connectivity of the corresponding Tanner graph. The considerably smaller size of the resulting tree, which follows from the smaller variable and check node degrees and from the less chance of invoking the pivoting rule, favors the memory-usage-intensive SSE algorithm significantly. As can be seen in Fig. 6(c), the composite approach C-UB possesses the greatest advantage over simple UBs that are not tight in multiplicity. The C-UB and the LB again tightly bracket the asymptotic performance.

*2) A (3,6) LDPC Code with $n = 72$:* The UB, the C-UB, the MC-S, and the tight LB are applied to bits 41, 25, and 60, as plotted in Fig. 7 and the complete statistics of all 72 bits are included in TABLE I(b). Almost all asymptotic orders and most multiplicities can be captured by the UB with only two exception bits. Both of the exception bits are of order 8, which is computed by the composite SSE for each bit respectively.

*3) A (3,6) LDPC Codes with $n = 144$:* Complete statistics of all 144 bits are presented in TABLE I(c), and we start to see many examples (101 out of 144 bits) in which the simple UB is not able to capture the asymptotic order and we have to resort to the C-UB for tighter results. It is worth noting that even the simple UB is able to identify some bits with BER of order 9, which requires $\binom{143}{8} \approx 3.55 \times 10^{12}$ trials if a brute force search is employed.[6]

### C. The Error-Floor Optimized LDPC Codes

For the randomly constructed LDPC codes in the previous examples, the frame-wise minimal stopping distances range from 2 to 4, which does not present any challenge for the proposed frame-wise SSE. That is the reason why we deliberately omit the discussion of the frame-wise minimal stopping set exhaustion in the previous examples and focus mainly on using the SSE as a bit-wise exhaustion or a bit-wise upper bound. For the following, we consider the regular and irregular codes optimized for the error-floor performance and having minimal stopping distances 8–13. The SSE will be used as the exhaustion algorithm and to provide exhaustive lists of minimal stopping sets. The trapping set exhaustion algorithm $k$-TSE will be applied to different codes of length $\approx 500$ as well. This range of codeword lengths is large enough to be of great practical importance and is beyond the reach of any brute-force search algorithm. On the other hand, it is still short enough for the proposed SSE and $k$-TSE algorithms to construct exhaustive lists of dominating error-prone patterns.

[6]For the bit-wise brute force search, since the target bit is always included, we need only to search through $\binom{143}{8}$ possibilities instead of $\binom{144}{9}$.

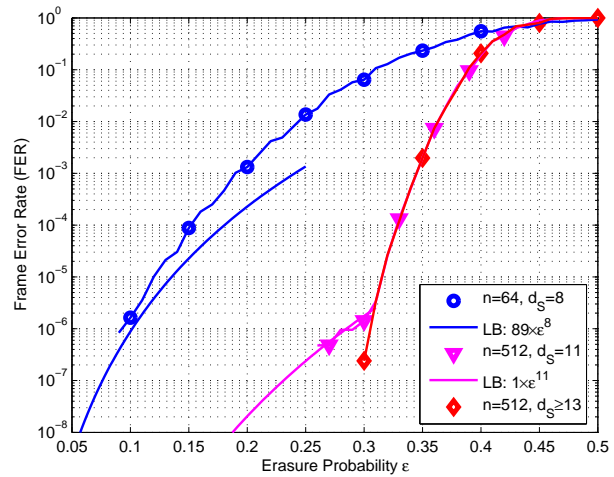

Fig. 8. Comparisons among three different regular (3,6) codes in the following order. A short $n = 64$ code with minimal stopping distance $d_S$ optimized to 8 and multiplicity 89 confirmed by the stopping set exhaustion. A randomly chosen $n = 512$ code with $d_S = 11$ and multiplicity 1 determined by the stopping set exhaustion. An optimized $n = 512$ code with $d_S \geq 13$. For the last code, one stopping set of size 24 is identified during the Monte-Carlo simulation.

The codes discussed herein are optimized by the code annealing method, presented in a companion paper [18], that uses the exhaustive lists of error-prone patterns generated by SSE and $k$-TSE as the objective functions for code optimization. For all the Monte-Carlo simulation results, 100 error events are observed for each simulation point.

#### Regular (3,6) LDPC Codes on BECs

*Experiment 1: An optimized (3,6) LDPC code with $n = 64$.* Using the exhaustive list as the objective function, its minimal stopping distance has been optimized to 8. Even for codes of this small size, the brute-force search requires $\binom{64}{8} \approx 4.43 \times 10^9$ computations, which can hardly be achieved by software simulations. All 89 minimal stopping sets are identified by SSE, and the exhaustive list provides a tight lower bound $89 \times \epsilon^8$ for the FER. The corresponding Monte-Carlo simulation results and the tight lower bound are plotted in Fig. 8. For comparison, the girth for any (3,6) code of length $n = 64$ is upper bounded by 6, which counts both variable and check nodes, while the minimal stopping set of this code consists of 8 variable nodes. This is a concrete example showing that even with a limited girth size (3 variable nodes), one can still construct codes with much larger minimal stopping distances.

*Experiment 2: A randomly chosen (3,6) LDPC code with*

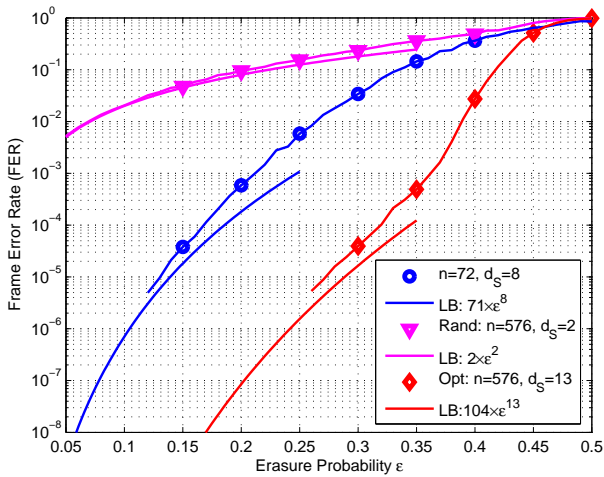

Fig. 9. Comparisons among three different irregular $(\lambda_1(x), \rho_1(x))$ codes with their minimal stopping distances $d_S$ and multiplicity $m_S$ determined by the minimal stopping set exhaustion. A short $n = 72$ code with $d_S$ optimized to 8 and $m_S = 71$. A randomly chosen $n = 576$ code, denoted by "Rand," with $d_S = 2$ and $m_S = 2$ has the poorest FER performance due to the existence of small stopping sets. In contrast with "Rand," the error floor of the optimized $n = 576$ code with $d_S = 13$ and $m_S = 104$, denoted by "Opt," is dramatically improved over its random counterpart. Tight lower bounds derived from the exhaustive lists of minimal stopping sets are plotted for comparison.

$n = 512$. It is well-known that regular (3,6) codes generally have superior error-floor performance with random construction, which is evidenced by this example. We arbitrarily choose one realization from the random code ensemble. This $n = 512$ (3,6) code has minimal stopping distance $d_S = 11$ with multiplicity 1. The SSE is able to identify the only one minimal stopping set, and the resulting tight lower bound coincides with Monte-Carlo simulation when $\epsilon \leq 0.3$; see Fig. 8. Using Monte-Carlo simulation, one can also identify the same minimal stopping set, but cannot claim with 100% certainty that there is no other stopping sets of size 11, which is a major distinction between the proposed SSE algorithm and any randomized enumeration algorithm, e.g. the error impulse methods.

*Experiment 3: An optimized (3,6) LDPC Code with $n = 512$.* An optimized (3,6) code with $n = 512$ is constructed by using the exhaustive list of stopping sets as the objective function. All candidates of sizes $\leq 12$ are exhausted and none of them is a stopping set. The minimal stopping distance must be $\geq 13$. One stopping set of size 24 is observed during the Monte-Carlo simulation, which gives us an upper bound on the actual minimal stopping distance. As can be seen in Fig. 8, for this optimized code with minimal stopping distance $\geq 13$, no error floor can be observed by software simulations until FER $= 10^{-7}$.

*Irregular LDPC Codes on BECs*

A more interesting subject is irregular LDPC codes with degree distributions optimized for threshold performance [11], which turns out to be the most computationally friendly instance for the proposed SSE algorithm. For rate-1/2 irregular codes of length 512, the searchable range $t$ can be extended to 13, which is +2 better than its regular code counterpart. Using the conventional polynomial representation of degree distributions [11], we consider codes of the following optimized variable and check node degree distributions:

$$\lambda_1(x) = 0.41667x + 0.16667x^2 + 0.41667x^5$$
$$\rho_1(x) = x^5,$$

which has asymptotic erasure probability threshold $\epsilon^* \approx 0.4775$ [53]. Codes of different codeword lengths and construction methods are discussed as follows and their results are illustrated in Fig. 9. For comparison, to exhaust all stopping sets of size 13 for a $n = 576$ code requires $\binom{576}{13} \approx 1.08 \times 10^{26}$ trials if a brute force approach is adopted.

*Experiment 4: An optimized $n = 72$ irregular LDPC code.* Using the above $(\lambda_1, \rho_1)$ degree distributions, an irregular code is constructed with optimized minimal stopping distance $d_S = 8$ as described in [18], and all of the 71 minimal stopping sets can be identified. Both codes in Experiments 1 and 4 are of minimal stopping distance 8 and of similar sizes, which show that although there are inherently many "bad" irregular codes, the error floor of "good" irregular codes is comparable to that of good regular codes, provided proper optimization is performed using the exhaustive list of minimal stopping sets as the objective functions.

*Experiment 5: An optimized $n = 576$ irregular LDPC code.* Using the same $(\lambda_1, \rho_1)$ degree distributions with $n = 576$, one can construct an irregular code with minimal stopping distance 13 [18]. All 104 minimal stopping sets have been identified by the SSE algorithm and a tight lower bound is provided. For comparison, we also plot the FER curve of a typical irregular code of the same degree distributions $(\lambda_1, \rho_1)$. At $\epsilon = 0.3$, an improvement of a factor 10,000 is reported when compared to the typical code performance, which again demonstrates the benefit of directly using the stopping set exhaustion list as the objective function.

*Irregular LDPC Codes on Gaussian Channels*

We focus our trapping set discussions on binary-input additive white Gaussian noise channels (BiAWGNCs). Namely, the observation vector $\mathbf{y} = (2\mathbf{x} - \mathbf{1}) + \sigma\mathbf{n}$, where $\mathbf{1}$ is an all-one vector, $\mathbf{n}$ is a standard Gaussian vector with unity covariance matrix, and $1/\sigma^2$ is the signal to noise (power) ratio. Only irregular LDPC codes will be considered and their common degree distributions, optimized for threshold performance, are as follows:

$$\lambda_2(x) = 0.31961x + 0.27603x^2 + 0.01453x^5 + 0.38983x^6$$
$$\rho_2(x) = 0.50847x^5 + 0.49153x^6,$$

with an asymptotic threshold $\sigma^* \approx 0.937$ [53]. All the minimal stopping sets or $k$-out trapping sets are identified by the SSE and $k$-TSE discussed previously. The corresponding Monte-Carlo simulation results are in Fig. 10.

*Experiment 6:* "Rand" is an $n = 512$ code arbitrarily chosen from the $(\lambda_2, \rho_2)$ irregular code ensemble with an additional constraint that there is no multiple edge in the corresponding bipartite Tanner graph. Without any graph optimization, the performance is not impressive for irregular codes of this short length. To further pinpoint the cause of the bad performance, its minimal $k$-out trapping sets are exhausted for $k=0,1$ (the 0-out trapping sets correspond to the stopping sets), and their (min_size,multiplicity) pairs are $(2,1)$ and $(2,8)$ respectively. Its short minimal trapping distances explain the poor performance of this code.

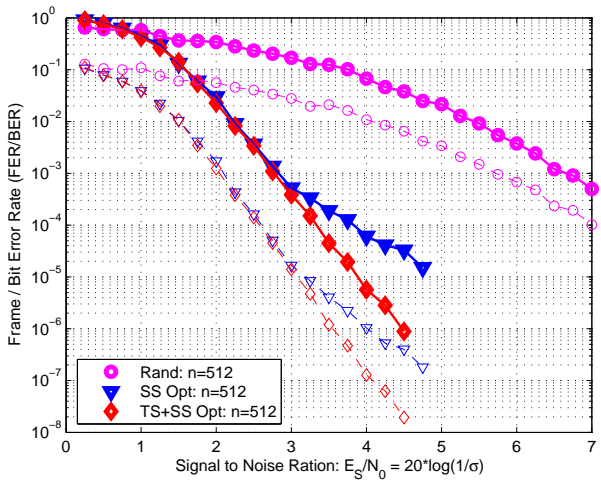

Fig. 10. Monte-Carlo simulations for irregular $(\lambda_2(x), \rho_2(x))$, $n=512$ codes with different constructions: "Rand," "SS Opt," and "TS+SS Opt" as described in Experiments 6, 7, and 8 respectively. Their corresponding (min_size,multiplicity) of the $k$-out trapping sets for $k=0,1$ respectively are: "Rand" $(2,1)$, $(2,8)$; "SS Opt" $(13,40)$, $(5,4)$; "TS+SS Opt" $(11,12)$, $(10,24)$. 80 decoding iterations are performed and 100 error events are observed for each simulation point.

*Experiment 7:* "SS Opt" is an $n=512$ code with degree distributions $(\lambda_2, \rho_2)$ but optimized for maximizing the minimal stopping distance by the code annealing method in [18]. Its corresponding (min_size,multiplicity) of the $k$-out trapping sets for $k=0,1$ are $(13,40)$ and $(5,4)$ respectively. This example confirms the phenomenon that codes optimized for BECs are generally good for other channels as well since the minimal 1-out trapping distance has been improved to 5 as a byproduct of optimizing the minimal stopping distance. This correlation between the BEC performance and the performance on other channel models has been used to develop several design heuristics for general channels, including partial elimination of stopping sets [36].

*Experiment 8:* "TS+SS Opt" is an $n=512$ code with degree distributions $(\lambda_2, \rho_2)$ but optimized for maximizing both the minimal 1-out trapping distance and the minimal stopping distance (the minimal 0-out trapping distance) by the code annealing method. Its corresponding (min_size,multiplicity) of the $k$-out trapping sets for $k=0,1$ are $(11,12)$ and $(10,24)$ respectively. Namely, for any set of contaminated bits of size $\leq 9$, there are at least two extrinsic messages that might break this error pattern. The BEC performance of "TS+SS Opt" is not as good as "SS Opt," since the minimal stopping distance is reduced from 13 to 11. However, the minimal 1-out trapping distance optimization further brings down the error floor for the non-erasure BiAWGNCs. No error floor is observed until FER=$10^{-6}$. The example demonstrates the benefit of using direct objective functions during code optimization in a channel-by-channel basis. The minimal trapping distance optimization is feasible due to the efficient minimal trapping set exhaustion Algorithm 4.

*Remark 1:* We did not exhaust the minimal 2-out trapping sets since there are many degree 2 nodes in these codes and each of them is by definition a 2-out trapping set.

*Remark 2:* When applied to codes of larger sizes, e.g., $n \geq 1000$, the searchable range $t$ of the SSE and $k$-TSE also increases when compared to $n \approx 500$ codes due to the larger girth and the more tree-like structure for longer codes. Nevertheless, the minimal stopping / trapping distances for longer codes generally grow faster than the searchable range $t$ and fewer tight results can be obtained for the error-floor optimized regular/irregular codes of length $n \geq 1000$.

### D. Algebraically Constructed Codes

We apply the exhaustion algorithms to the following well-studied codes: the (155,64,20) Tanner code [29], the Ramanujan-Margulis (2184,1092) code with $q=13$ and $p=5$ [46], and the (672,336) Margulis code with $p=7$ [47]. Since all these codes are regular codes of minimal variable node degree 3, the 2-out trapping set exhaustion is performed in addition to the stopping set and the 1-out trapping set exhaustion.

*1) The (155,64,20) Tanner Code:* The minimal Hamming distance of this code is known to be 20, which is computed by MAGMA, a software suite taking advantage of the corresponding algebraic structure [54]. One can easily locate a stopping set of size 18 by Monte-Carlo simulation, which serves as an upper bound of the minimal stopping distance. The SSE and $k$-TSE are then performed to gather new results regarding different types of error-prone patterns of this code.

For stopping sets (0-out trapping sets), all candidates of size $\leq 12$ have been exhausted and none of them is a stopping set, which results in a lower bound 13 on the minimal stopping distance. For 1-out trapping sets, all candidates of size $\leq 11$ have been exhausted and none of them is a 1-out trapping set, which results a lower bound 12 on the minimal 1-out trapping distance. For 2-out trapping sets, we are able to exhaustively locate all 465 minimal 2-out trapping sets, which are of size 8. All 465 minimal 2-out trapping sets can be obtained from the following 5 representatives by the automorphisms discussed in [29].

$$7, 17, 19, 33, 66, 76, 128, 140$$
$$7, 31, 33, 37, 44, 65, 100, 120$$
$$1, 19, 63, 66, 105, 118, 121, 140$$
$$44, 61, 65, 73, 87, 98, 137, 146$$
$$31, 32, 37, 94, 100, 142, 147, 148.$$

Recently, the instanton analysis in [16], [9] identified some dominating error patterns, termed as instantons in their paper. Each of the reported instantons contains one minimal 2-out trapping set as a substructure.[7] These results again confirm

---

[7] In [16], [9], the constituting permutation matrices are cyclically shifted to the right, which results in different indexing than the original construction in which the permutation matrices are cyclically shifted to the left [29]. An index remapping is necessary for direct comparison.

that the dominating error patterns are generally $k$-out trapping sets with small $k$.

*2) The Ramanujan-Margulis (2184,1092) Code w. $q = 13, p = 5$:* MacKay *et al.* [7] first pointed out there is a codeword of Hamming weight 14 for the Ramanujan-Margulis (2184,1092) code with $q = 13, p = 5$. By the automorphism of the code, there are at least 1,092 codewords of such weight. By a refinement of the error impulse method, Hu *et al.* [32] also identified the same 1,092 codewords for the Ramanujan-Margulis (2184,1092) code. The remaining question is whether there is any other codeword of equal or smaller Hamming weight. There are at least two approaches to answering this question: one is to use the algebraic structure of the Ramanujan-Margulis code and mathematically prove that there is no other codeword of equal or smaller weights. The second approach is by the exhaustive search, which requires $6.2 \times 10^{35}$ trials if by a brute force search but is feasible by the SSE algorithms.

Our stopping set exhaustion algorithm shows that there are "only" 1,092 stopping sets of size 14 and there is no stopping set of smaller size. Since any codeword must be a stopping set, the minimal Hamming distance of the Ramanujan-Margulis (2184,1092) code must be 14 and its multiplicity must be 1,092. What MacKay *et al.* and Hu *et al.* found are indeed the minimal codewords.

We also exhaust its $k$-out trapping sets for $k = 1, 2$ respectively. All candidates of 1-out trapping sets of size $\leq 12$ and all candidates of 2-out trapping sets of size $\leq 9$ are exhausted and none of them is a trapping set, which results in lower bounds of 13 and 10 on the minimal 1-out and 2-out trapping distances, respectively. The lack of trapping sets of small sizes explains why the dominating error event of the Ramanujan-Margulis (2184,1092) code concerns the small codewords instead of the small near-codewords.

*3) The (672,336) Margulis code with $p = 7$ [47]:* It is known that this code has a codeword of length 16 [9]. The SSE and $k$-TSE are again performed to gather new results regarding the error-prone patterns of this code.

For stopping sets, the SSE shows that there is no stopping set of size $\leq 13$. For $k$-out trapping sets, there is no 1-out trapping set of size $\leq 12$ and there is no 2-out trapping of size $\leq 9$. Since no exhaustive list of any type of minimal error-patterns can be obtained, our results serve only as lower bounds on the sizes of minimal error-prone patterns. For comparison, all stopping sets observed by the Monte-Carlo simulation are of size $\geq 16$ and no tighter upper bound can be obtained for the minimal stopping distance. Combined with our results, the minimal stopping and the Hamming distances of the (672,336) Margulis code are between 14–16.

Results for the three algebraically constructed codes are summarized in TABLE II.

*Note:* The SSE and $k$-TSE are directly applied to these codes without taking advantage of their algebraic structure except for the automorphism, which is used only in the statement that results of the bit-wise search is sufficient to determine the corresponding results for the entire frame. The frame-wise search can thus be simplified to the more efficient bit-wise search.

|  | Tanner(155,64) | R-M(2184,1092) | Marg(672,336) |
|---|---|---|---|
| Hamming Dist. | | | |
| by enum. | $(20, ?)^\dagger$ | $(\leq 14, ?)$ | $(\leq 16, ?)$ |
| by exhaust'n | $(\geq 13, ?)$ | $(14, 1092)$ | $(\geq 14, ?)$ |
| Stopping Dist. | | | |
| by enum. | $(\leq 18, ?)$ | $(\leq 14, ?)$ | $(\leq 16, ?)$ |
| by exhaust'n | $(\geq 13, ?)$ | $(14, 1092)$ | $(\geq 14, ?)$ |
| 1-Out Trap. Dist. | | | |
| by exhaust'n | $(\geq 12, ?)$ | $(\geq 13, ?)$ | $(\geq 13, ?)$ |
| 2-Out Trap. Dist. | | | |
| by exhaust'n | $(8, 465)$ | $(\geq 10, ?)$ | $(\geq 10, ?)$ |

TABLE II

Summary of bounding the minimal Hamming, stopping, 1-out trapping, and 2-out trapping distances for the Tanner (155,64,20) code, the Ramanujan-Margulis (2184,1092) code with $q = 13, p = 5$, and the Margulis (672,236) code with $p = 7$. The pair $(d, m)$ representing the minimal distance $d$ (or the range of the minimal distance) and how many codewords/stopping sets/trapping sets are of weight $d$. Upper bounds are obtained from enumerations, while lower bounds are from the exhaustion algorithm provided herein. † is obtained using computer search MAGMA [29].

## VIII. CONCLUSION & FUTURE RESEARCH

The problem of exhausting error-prone patterns for arbitrary LDPC codes with finite length has been thoroughly discussed, and efficient exhaustion algorithms have been devised for obtaining both the minimal stopping sets for binary erasure channels and the minimal trapping sets for general channels. The algorithms, based on the decoding tree of the iterative decoder, are equivalent to a narrowing search of potential error prone patterns. The optimality of the algorithms has been proven and several improvements have been made to further improve the efficiency, including a composite approach combining the divide-and-conquer strategy of the brute-force search and the proposed algorithm. All the proofs and derivations are based on the proposed Boolean expression framework for iterative decoding functions over binary erasure channels. Extensive numerical experiments have been conducted on different codes, including randomly constructed codes, error-floor optimized codes, and algebraically constructed codes. For rate-1/2 codes of short practical lengths $n \approx 500$, both the minimal stopping sets and the minimal trapping sets can be listed exhaustively. The exhaustive list of error-prone patterns finds applications in code behavior analysis, upper and lower bounding code performance, and finite code optimization for improved error-floor performance.

The NP-completeness of the trapping set exhaustion algorithm has also been established, which demonstrates the inherent hardness of this problem. Implying that there is little chance that an efficient algorithm for exhaustively enumerating small stopping / trapping sets exists, the NP-completeness argument is genuinely an asymptotic worst-case analysis, which does not preclude the existence of efficient algorithms for codes of short lengths. The algorithms discussed in this work serve as the first successful step toward addressing the seemingly intractable problem of exhausting error-prone patterns for LDPC codes of practical lengths.

There are many possible approaches to enhancing the efficiency of the proposed algorithms or to designing better algorithms for even longer codes. We conclude this paper by providing a short, non-comprehensive list of potential improvements.

1) Several parts of the existing algorithms can be further improved, including but not limited to a better design of the leaf-finding module, as discussed previously, and the partitioning events for the composite approach. The effects of different partitioning events on the efficiency of the algorithms can easily be as large as a factor of 100. With considerable amount of design freedom over the partitioning events, we believe the full power of the composite approach is not utilized. One possible direction is to use the information from the Monte-Carlo simulation as a design guideline for partitioning events.
2) The efficiency of the current algorithms is sufficient for codes of short lengths $n \approx 500$. Better implementations with different data structures, machine-optimized software codes, or even a hardware implementation of the algorithms could further improve the applicable range to $n \approx 1000$ or larger, which will have even broader impact on finite length LDPC code analysis and design.
3) The minimal trapping sets are the error-floor determining factor for general channels, and the sizes of interest are generally less than 15, as opposed to the minimal stopping distances of interest, which are generally 15–30. The above reasons make the exhaustion of trapping sets an especially appealing problem due to its practical importance and its much smaller search range. In this work, the minimal trapping set exhaustion problem is solved by the reduction to the minimal stopping set exhaustion, and significant efficiency loss is incurred. Directly designing algorithms for the minimal trapping set exhaustion should recover the performance loss and extend the applications to codes of medium length $n \approx 1000$.

# APPENDIX I
# DETAILED DESCRIPTIONS OF Algorithm 3

Some technical details of Algorithm 3 are discussed herein.

*How to grow a leaf from the Tanner graph?*

Since every time a variable node $v_k$ is added to $\mathcal{T}$, all its immediate check node children are added to $\mathcal{T}$ as well,[8] a "leaf" in $\mathcal{T}$ actually refers to a small sub-tree rooted at $v_k$, consisting of $v_k$ and all its immediate children $c_j$. Therefore, all possible "positions" of growing "leaves" are check nodes. Suppose we would like to grow another leaf at check node $c_{j_0}$. From $c_{j_0}$'s immediate parent $v_{i_0}$, one can look up the directed $(c_{j_0}, v_{i_0})$ edge in the Tanner graph, and consider its incoming extrinsic message flows $(v_i, c_{j_0})$ such that $i \neq i_0$. If one $v_i$ is chosen and added to $\mathcal{T}$ through $c_{j_0}$, then the immediate children of the newly-added $v_i$ refer to those $c_j \neq c_{j_0}$ connected to $v_i$ in the Tanner graph. There are cases in which some $v_i$'s have already been added to $c_{j_0}$ while some $v_i$'s have not. Then the next leaf to grow has to be chosen

[8]The only exception is during the finishing step Line 16.

from those $v_i$'s that have not been previously added to check node $c_{j_0}$.

*Active and Inactive Variable Nodes*

A special notation in Algorithm 3 is that each variable node $v_k$ has been assigned a status either active or inactive. The reasons behind this notation are as follows. Unlike the Tanner graph, in which each variable node appears exactly once, in the infinite-sized decoding tree, the same variable node may show up infinitely many times. Therefore, the term "a variable node $v_k$" has two different meanings in that it either refers to "the node" in the decoding tree $\mathcal{T}$ associated with an AND operation or refers to the corresponding "variable" in the Tanner graph accepting observation $y_k$. The status of a variable node is to emphasize the link between the free observation $y_k$ in the Tanner graph and a node located in the decoding tree. Each active node serves two purposes: taking a free observation $y$ as its input and acting as an AND operation in the iterative decoding function. For an inactive node, no free input observation is taken and it acts only as an AND operation.

*Note:* Although originally not associated with any $v_k$ in the Tanner graph, an auxiliary variable node can still be active and has free observation $y_k$ as illustrated in the root of the left subtree in Fig. 4(b). Those auxiliary nodes have to be considered in Line 7 and therefore, the node $v'_k$ may be an auxiliary tree node and have no direct connection to any node in the Tanner graph.

yca$(v_k)$ *and* $v'_k$

For any two nodes in a tree, their youngest common ancestor can be uniquely defined by finding all the common ancestors of this pair of nodes and then selecting the youngest of them. For any existing active node with free observation $y_k$, we can uniquely determine the youngest common ancestor between the newly-added leaf and the existing node. In Line 8, yca$(v_k)$ refers to the youngest among the set of all youngest common ancestors generated by different pairs of $v_k$ versus existing active variable nodes, and $v'_k$ is the corresponding active node with free observation $y_k$ resulting yca$(v_k)$ jointly with $v_k$. As observant readers might notice, yca$(v_k)$ is unique but $v'_k$ may not be. Algorithm 3 works for any choice of $v'_k$.

The pivoting rule depicted in Fig. 4 relies on the assumption that among all messages coming from the children of yca$(v_k)$, there are exactly two messages, denoted by $g(\mathbf{y})$ and $h(\mathbf{y})$, sharing a single common determining variable $y_k$. This assumption will be proved as a byproduct of proving the correctness of Algorithm 3.

*Converting a tree $g(\mathbf{y})$ to its conditional version $g(\mathbf{y}_0^{k-1} y_k \mathbf{y}_{k+1}^n)$ for a fixed $y_k \in \{0, 1\}$.*

Any subtree corresponding to a function $g(\mathbf{y})$ can be converted to a conditional subtree $g(\mathbf{y}_0^{k-1} 0 \mathbf{y}_{k+1}^n)$ as follows. Locate *all* active variable nodes having free observation $y_k$. Change their status to inactive by hardwiring their input observation $y_k$ to 0. Although in Fig. 4(a) only a single square is used to represent $y_k$, the conditional function $g(\mathbf{y}_0^{k-1} 0 \mathbf{y}_{k+1}^n)$ can be obtained only by hardwiring all free observations $y_k$ in the corresponding subtree since $y_k$ is generally involved in more than one product form. The subtree conditioned on $y_k = 1$ can be obtained similarly.

*Evaluating* $\text{UB}_i$

The evaluation of $\text{UB}_i$ proceeds iteratively based on the formulas in Rule 0. For the case in which an observation $y_k$ is hardwired to zero/one, we substitute $\mathsf{E}\{y_k\} = 0$ or $\mathsf{E}\{y_k\} = 1$ into the formula of the variable node function $\mathsf{E}\{f_v(\mathbf{y})\}$. Note: the formulas in Rule 0 become pure arithmetic operations computing an upper bound of the expected value and lose their interpretations as computing the expectations. A more accurate notation for the iterative upper bound computation should be

$$\begin{aligned}\text{UB}_{\mathsf{E}\{f_v(\mathbf{y})\}} &= \text{UB}_{\mathsf{E}\{g(\mathbf{y})\}} \text{UB}_{\mathsf{E}\{h(\mathbf{y})\}} \\ \text{UB}_{\mathsf{E}\{f_c(\mathbf{y})\}} &= \text{UB}_{\mathsf{E}\{g(\mathbf{y})\}} + \text{UB}_{\mathsf{E}\{h(\mathbf{y})\}} \\ &\quad - \text{UB}_{\mathsf{E}\{g(\mathbf{y})\}} \text{UB}_{\mathsf{E}\{h(\mathbf{y})\}},\end{aligned}$$

instead of using the expectation operator.

## APPENDIX II
## A PROOF OF THE CORRECTNESS OF Algorithm 3

The proof of the correctness of Algorithm 3 consists of the following two steps. For any $\mathcal{T}$ constructed by Algorithm 3 after $t$ iterations, let $f_{\mathcal{T},t}(\mathbf{y})$ denote the corresponding iterative decoding function. We will show that (i) $\mathsf{E}\{f_{\mathcal{T},t}(\mathbf{y})\}$ is no larger than the $\text{UB}_i$ computed by Algorithm 3, and (ii) the decoding function $f_i(\mathbf{y}) \leq f_{\mathcal{T},t}(\mathbf{y})$ for all $\mathbf{y}$. Since (ii) is a restatement of *Theorem 3*, we will prove (i) in this appendix by the following two propositions and leave (ii) to APPENDIX III.

*Proposition 3:* In the tree $\mathcal{T}$ generated by Algorithm 3, all messages entering the same variable nodes must be independent. Namely, if $g(\mathbf{y})$ and $h(\mathbf{y})$ enter the same variable node, then they share no common determining variable.

*Proof:* An equivalent statement of the above proposition is that the youngest common ancestor of any two active variable nodes with the same free observation must be a check node. A simple proof of the equivalence is as follows. Suppose there exist two distinct, dependent messages $g(\mathbf{y})$ and $h(\mathbf{y})$ entering the same variable node $v$. Then $g(\mathbf{y})$ and $h(\mathbf{y})$ share at least one common determining variable, denoted by $y_k$, and there must be active variable nodes with free observation $y_k$ in each of the subtrees of $g(\mathbf{y})$ and $h(\mathbf{y})$. Pick one active variable node from each subtree. The youngest common ancestor of the selected pair of active variable nodes is thus variable node $v$, which contradicts the assumption that the youngest common ancestor of any pair of active variable nodes with the same free observation is always a check node. The equivalence is obtained.

We then prove by induction that all youngest common ancestors, if they exist, are check nodes. During the initialization, there is only one active variable $v_i$ with free observation $y_i$ in $\mathcal{T}$, and the above statement obviously holds. Suppose the statement holds after $t$ iterations of the REPEAT–UNTIL loop in Line 3. For the $(t + 1)$-th iteration of adding $v_k$ into $\mathcal{T}$, consider two cases: $\text{yca}(v_k)$, the youngest of all common ancestors, is a check node or not. Using the same notation as in Algorithm 3, we use $v'_k$ to denote the existing active variable node resulting in $\text{yca}(v_k)$ jointly with $v_k$. For the case in which $\text{yca}(v_k)$ is a check node, no pivoting is involved so one needs only to show that after adding $v_k$ there is no other active node $v''_k$ with free observation $y_k$ such that the youngest common ancestor of $v_k$ and $v''_k$ is a variable node $v$. We prove this statement again by contradiction. Suppose there exists such a $v''_k$. Since $\text{yca}(v_k)$ is the youngest of all common ancestors and is a check node, $\text{yca}(v_k)$ must be a strict descendent of $v$. Therefore, the youngest common ancestor between $v'_k$ and $v''_k$ must also be $v$, which contradicts the assumption that after $t$-iterations all youngest common ancestors of pairs of active variable nodes with the same free observation are check nodes.

Before proving the case in which $\text{yca}(v_k)$ is a variable node, we first show that if $\text{yca}(v_k)$ is a variable node, before pivoting, there are exactly two messages $g(\mathbf{y})$ and $h(\mathbf{y})$ entering $\text{yca}(v_k)$ such that they share a single common determining variable $y_k$. A short argument is as follows. Since $\text{yca}(v_k)$ is a youngest common ancestor of nodes $v_k$ and $v'_k$, there must be two completely disjoint paths from $\text{yca}(v_k)$ to $v_k$ and $v'_k$ respectively, which implies that $v_k$ and $v'_k$ must be attributed to different messages entering $\text{yca}(v_k)$. Therefore, there are at least two incoming messages having $y_k$ as their determining variables. Suppose there is another message $g'(\mathbf{y})$ entering $\text{yca}(v_k)$ which also has $y_k$ as its determining variable. Then there must be an active variable node $v''_k$ in the subtree corresponding to $g'(\mathbf{y})$ that also accepts free observation $y_k$. The youngest common ancestor of nodes $v'_k$ and $v''_k$ must be $\text{yca}(v_k)$, which contradicts the assumption that all youngest common ancestors must be check nodes after $t$-iterations.

To show that after pivoting (i.e., when $\text{yca}(v_k)$ is a variable node) the same statement holds, we first note that after pivoting, all active nodes in the subtree rooted at $\text{yca}(v_k)$ with free observation $y_k$ are deactivated, with the only exception being the root of the left subtree, see Fig. 4 for illustration. The active variable node $v_k$ is thus referring to two different nodes before and after pivoting. Before pivoting, $v_k$ refers to the newly added leaf. After pivoting, $v_k$ refers to the root of the left subtree. Suppose after pivoting there exists another active variable node $v''_k$ with free observation $y_k$ such that the youngest common ancestor of $v_k$ and $v''_k$ is a variable node $v$. As shown previously, before pivoting, there are exactly two messages $g(\mathbf{y})$ and $h(\mathbf{y})$ having $y_k$ as their determining variables. So before pivoting, all active variable nodes in the subtree rooted at $\text{yca}(v_k)$ with free observation $y_k$ are within the only two subtrees corresponding to $g(\mathbf{y})$ and $h(\mathbf{y})$. After pivoting, all those active nodes will be deactivated with the only exception being $v_k$. Therefore, $v''_k$, as an active variable node after pivoting, cannot be a descendent of $\text{yca}(v_k)$, which implies that $v$ must be a strict ancestor of $\text{yca}(v_k)$.

The fact that after pivoting $v''_k$ is not a descendent of $\text{yca}(v_k)$ and $v$ is a strict ancestor of $\text{yca}(v_k)$ implies that before pivoting, the youngest common ancestor of active nodes $v''_k$ and $v'_k$ is also $v$, which contradicts the assumption that after $t$-iterations all youngest common ancestors of pairs of active variable nodes with the same free observation are check nodes.

Another possibility is that after pivoting with respect to $v_k$, there exist two active nodes $v_\chi$ and $v'_\chi$, both of whom are accepting free observation $y_\chi \neq y_k$ and the youngest common ancestor of this pair is a variable node. We consider the following four cases regarding to the tree configuration after pivoting: (a) neither $v_\chi$ nor $v'_\chi$ is a descendent of $\mathsf{yca}(v_k)$, (b) one of them, say $v_\chi$, is a descendent of $\mathsf{yca}(v_k)$ while the other is not, (c) both of them are in the same left (or right) subtree, and (d) $v_\chi$ and $v'_\chi$ are in the left and the right subtrees respectively. For case (a), since the pivoting rule only affects the descendents of $\mathsf{yca}(v_k)$, $v_\chi$ and $v'_\chi$ are intact during pivoting and their youngest common ancestor must be a check node due to the induction assumption. For case (b), the youngest common ancestor of $v_\chi$ and $v'_\chi$ after pivoting must be an ancestor of $\mathsf{yca}(v_k)$ since $v_\chi$ is a descendent of $\mathsf{yca}(v_k)$ but $v'_\chi$ is not. Since $v_\chi$ must be obtained from an active variable node during the duplicating step of pivoting, we use $v''_\chi$ to denote the original active variable node before pivoting. Then the youngest common ancestors between $(v_\chi, v'_\chi)$ and $(v''_\chi, v'_\chi)$ coincide and must be a check node. For case (c), since $v_\chi$ and $v'_\chi$ are in the same left (or right) subtree, their youngest common ancestor must also be in the same left (or right) subtree. As a result, the relative locations among $v_\chi$, $v'_\chi$, and their ancestor must be preserved during pivoting. Since before pivoting the youngest common ancestor between any two active nodes accepting free observation $y_\chi$ must be a check node, the youngest common ancestor of $v_\chi$ and $v'_\chi$ must be a check node as well. For case (d), the youngest common ancestor of $v_\chi$ and $v'_\chi$ is the check node of degree 2 taking the two subtrees as its children.

From the above reasoning, after $(t+1)$ iterations, all youngest common ancestors of pairs of active variable nodes with the same free observation are check nodes. *Proposition 3* is thus proved by induction. ∎

*Proposition 4:* The $\mathrm{UB}_i$ computed in Algorithm 3 is an upper bound on $\mathsf{E}\{f_{\mathcal{T},t}(\mathbf{y})\}$. Furthermore, $\mathrm{UB}_i$ is of the same asymptotic order as that of $\mathsf{E}\{f_{\mathcal{T},t}(\mathbf{y})\}$.

*Proof:* By *Proposition 3*, all messages entering the variable nodes are independent, and one can use Rule 0 to compute the exact expectation of the outgoing message from a variable node. On the other hand, Rule 1 gives us an upper bound on the expectation of the outgoing message from a check node, namely,

$$\mathsf{E}\{f_c(\mathbf{y})\} \leq \mathrm{UB}_{\mathsf{E}\{f_c(\mathbf{y})\}}$$
$$= \mathsf{E}\{g(\mathbf{y})\} + \mathsf{E}\{h(\mathbf{y})\} - \mathsf{E}\{g(\mathbf{y})\}\mathsf{E}\{h(\mathbf{y})\}.$$

Using Rule 0 for $\mathsf{E}\{f_v(\mathbf{y})\}$ and Rule 1 for $\mathsf{E}\{f_c(\mathbf{y})\}$, we are able to compute their upper bounds assuming we know the expectations of all input messages, which are unfortunately not available and are the quantities to be bounded. Fortunately, both Rule 0 and Rule 1 are monotonically increasing functions with respect to the input expectations, and we can thus substitute all input expectations with their upper bounds and obtain the iterative upper bounding formula:

$$\mathrm{UB}_{\mathsf{E}\{f_v(\mathbf{y})\}} = \mathrm{UB}_{\mathsf{E}\{g(\mathbf{y})\}} \mathrm{UB}_{\mathsf{E}\{h(\mathbf{y})\}}$$
$$\mathrm{UB}_{\mathsf{E}\{f_c(\mathbf{y})\}} = \mathrm{UB}_{\mathsf{E}\{g(\mathbf{y})\}} + \mathrm{UB}_{\mathsf{E}\{h(\mathbf{y})\}}$$
$$\qquad - \mathrm{UB}_{\mathsf{E}\{g(\mathbf{y})\}} \mathrm{UB}_{\mathsf{E}\{h(\mathbf{y})\}},$$

which is as if we are computing the expectation value by blindly assuming all messages are independent. Since the only inequality involved is resulted from Rule 1, which induces no order loss, the $\mathrm{UB}_i$ is an upper bound of $\mathsf{E}\{f_{\mathcal{T},t}(\mathbf{y})\}$ tight in the asymptotic order. ∎

## Appendix III
### The Proof of *Theorem 3*

Since the collection of all stopping sets involving bit $x_i$ can be rewritten as $\{\mathbf{x} \subseteq \{x_1, \ldots, x_n\} : f_i(\mathbf{y_x}) = 1\}$ where $f_i(\mathbf{y})$ is the decoder for bit $x_i$, proving *Theorem 3* is equivalent to showing that $f_i(\mathbf{y}) \leq f_{\mathcal{T},t+1}(\mathbf{y}) \leq f_{\mathcal{T},t}(\mathbf{y})$ for all $\mathbf{y} \in \{0,1\}^n$, $t \in \mathbb{N}$. To better describe the tree-revealing argument used in the proof, we introduce the following lemma and notation.

For any infinite-sized tree $\mathcal{T}$ corresponding to an iterative decoding function $f_{\mathcal{T}}(\mathbf{y})$, a *localized* subtree of $\mathcal{T}$ is defined as a subtree $\mathcal{U}$ sharing the same root as $\mathcal{T}$. The function $f_{\mathcal{U}}(\mathbf{y})$ is computed by assuming all edges/messages outside of $\mathcal{U}$ are hardwired to 1. Namely, the iterative decoding $f_{\mathcal{U}}(\mathbf{y})$ is performed only within the localized subtree $\mathcal{U}$. By the channel degradation argument, we have the following self-explanatory lemma.

*Lemma 2:* If $\mathcal{U}$ is a localized subtree of $\mathcal{T}$, then $f_{\mathcal{U}}(\mathbf{y}) \leq f_{\mathcal{T}}(\mathbf{y})$ for all $\mathbf{y} \in \{0,1\}^n$.

*Proof of Theorem 3:* We use $\overline{\mathcal{T}}_{x_i}$ to denote the infinite-sized tree corresponding to the iterative decoder $f_i(\mathbf{y})$, and use $\mathcal{T}_t$ to denote the finite tree $\mathcal{T}$ of Algorithm 3 after $t$ iterations of the REPEAT–UNTIL loop where $t \in \{0, 1, \ldots\}$. We also define $\overline{\mathcal{T}}_0 = \overline{\mathcal{T}}_{x_i}$ to be the first entry of a series of infinite-sized trees. By noting that $\mathcal{T}_0$, containing $v_i$ and its immediate cheeck node children, is a localized subtree of $\overline{\mathcal{T}}_0$ and by *Lemma 2*, we have

$$f_{\mathcal{T},0}(\mathbf{y}) = f_{\mathcal{T}_0}(\mathbf{y})$$
$$\geq f_{\overline{\mathcal{T}}_0}(\mathbf{y}) = f_{\overline{\mathcal{T}}_{x_i}}(\mathbf{y}) = f_i(\mathbf{y}), \forall \mathbf{y} \in \{0,1\}^n.$$

We then consider the situation after the first iteration, which naturally consists of two cases corresponding to whether pivoting is performed or not during the first iteration. For the simpler case, in which $\mathsf{yca}(v_k)$ is a check node and no pivoting is performed, define $\overline{\mathcal{T}}_1 = \overline{\mathcal{T}}_0$ as the second entry of the series of infinite-sized trees. By observing that $\mathcal{T}_1$ is obtained from $\mathcal{T}_0$ by adding another leaf that was previously in $\overline{\mathcal{T}}_0$ but not in $\mathcal{T}_0$, we have that $\mathcal{T}_1$ is a localized subtree of $\overline{\mathcal{T}}_1 = \overline{\mathcal{T}}_0$, and $\mathcal{T}_0$ is a localized subtree of $\mathcal{T}_1$. Therefore,

$$f_{\mathcal{T},0}(\mathbf{y}) = f_{\mathcal{T}_0}(\mathbf{y})$$
$$\geq f_{\mathcal{T},1}(\mathbf{y}) = f_{\mathcal{T}_1}(\mathbf{y})$$
$$\geq f_{\overline{\mathcal{T}}_1}(\mathbf{y}) = f_{\overline{\mathcal{T}}_0}(\mathbf{y}) = f_i(\mathbf{y}), \quad \forall \mathbf{y} \in \{0,1\}^n. \quad (10)$$

The more interesting case is when a pivoting is performed. Let $\mathcal{T}_{1/2}$ denote the tree after adding another leaf but before pivoting. By Lemma 2, we have $f_{\mathcal{T},0}(\mathbf{y}) = f_{\mathcal{T}_0}(\mathbf{y}) \geq f_{\mathcal{T}_{1/2}}(\mathbf{y})$. Since what pivoting does is to transform the original tree $\mathcal{T}_{1/2}$ into its equivalent form $\mathcal{T}_1$, we have $f_{\mathcal{T},1}(\mathbf{y}) = f_{\mathcal{T}_1}(\mathbf{y}) = f_{\mathcal{T}_{1/2}}(\mathbf{y})$ for all possible $\mathbf{y}$. Furthermore, the same pivoting operation can be applied to the infinite-sized $\overline{\mathcal{T}}_0$ and we denote the end result as $\overline{\mathcal{T}}_1$. Similarly, we will have $f_{\overline{\mathcal{T}}_1}(\mathbf{y}) = f_{\overline{\mathcal{T}}_0}(\mathbf{y})$. By noting that $\mathcal{T}_1$ is again a localized subtree of $\overline{\mathcal{T}}_1$, (10) holds for the case in which pivoting is performed as well. The above completes the proof of the induction from $t = 0$ to $t = 1$. The induction from $t$-th iteration to $t+1$-th iteration follows analogously. The proof of *Theorem 3* is complete. ∎

## APPENDIX IV
## PROOF OF *Theorem 4*

*Theorem 4* is the culmination of all analyses of Algorithm 3, the proof of which needs the following concepts / notation in addition to those established in APPENDICES I through III.

*The Transcribed Tree $\mathcal{T}$*

Following the same notation as in APPENDIX III, let $\overline{\mathcal{T}}_{x_i}$ denote the infinite-sized tree corresponding to the iterative decoder $f_i(\mathbf{y})$. Denote a series of localized finite subtrees of $\overline{\mathcal{T}}_{x_i}$ by $\{\mathcal{T}_{x_i,0}, \ldots, \mathcal{T}_{x_i,s}, \ldots\}$ such that $\mathcal{T}_{x_i,0}$ contains only one variable node $v_i$ and its immediate children. $\mathcal{T}_{x_i,s+1}$ can be iteratively constructed from $\mathcal{T}_{x_i,s}$ by adding a single leaf to $\mathcal{T}_{x_i,s}$. Note: Unlike Algorithm 3, no pivoting is performed this time. $\mathcal{T}_{x_i,s}$ just includes more and more of $\overline{\mathcal{T}}_{x_i}$ as $s$ becomes large.

Let $\mathcal{T}_t$ denote the tree resulting from Algorithm 3 after $t$ iterations of the REPEAT-UNTIL loop in Line 3. The definition of the transcript tree is as follows.

*Definition 4:* For fixed $t$ and $s$, $\mathcal{T}_t$ is a tree transcribed from $\mathcal{T}_{x_i,s}$ if $f_{\mathcal{T}_t}(\mathbf{y}) = f_{\mathcal{T}_{x_i,s}}(\mathbf{y})$ for all possible $\mathbf{y}$.

Namely, a transcribed tree $\mathcal{T}_t$ is a tree of a different structure than $\mathcal{T}_{x_i,s}$ but is equivalent to $\mathcal{T}_{x_i,s}$ from the output value perspective.

*The Balanced Growing LF Module*

A balanced growing leaf finding (LF) module is defined as follows.

*Definition 5 (The Balanced Growing LF Module):* With the assumption that Algorithm 3 is combined with the pruning rules in Section V-A, the corresponding LF module is balanced growing with respect to the series $\{\mathcal{T}_{x_i,0}, \ldots, \mathcal{T}_{x_i,s}, \ldots\}$ if there exists an increasing list of time instants $t_1 < t_2 < \ldots < t_s < \ldots$, such that $\mathcal{T}_{t_s}$ is a transcribed tree of $\mathcal{T}_{x_i,s}$ for all $s \in \mathbb{N}$.

*Proposition 5:* The balanced growing LF module exists for any sequence $\{\mathcal{T}_{x_i,0}, \ldots, \mathcal{T}_{x_i,s}, \ldots\}$.

*Proof:* This proposition will be proved by explicit construction.

When $s = 0$, simply set $t_0 = 0$. Since $\mathcal{T}_{t_0}$ contains only variable node $v_i$ and its immediate children and is identical to the first entry $\mathcal{T}_{x_i,0}$, $\mathcal{T}_{t_0}$ is a transcribed tree of $\mathcal{T}_{x_i,0}$. *Proposition 5* holds for $s = 0$.

Suppose *Proposition 5* holds for the general $s$-th entry, i.e., $\mathcal{T}_{t_s}$ is a transcribed tree of $\mathcal{T}_{x_i,s}$. Suppose $\mathcal{T}_{x_i,s+1}$ is obtained from $\mathcal{T}_{x_i,s}$ by adding a leaf $v_k$. Using the notation of APPENDIX III, let $\overline{\mathcal{T}}_{t_s}$ denote the infinite-sized tree after $t_s$ iterations, of which $\mathcal{T}_{t_s}$ a localized subtree.

We first note that the pivoting operation involves duplicating trees, which results in a one-to-many relationship between nodes in the original tree and nodes in the pivoted tree. Due to the fact that the pivoting operation may be performed many times before arriving at $\mathcal{T}_{t_s}$, there may be more than one leaves in $\overline{\mathcal{T}}_{t_s}$ but not in $\mathcal{T}_{t_s}$ that correspond to the to-be-added leaf $v_k$ for $\mathcal{T}_s$. All these leaves are adjacent to $\mathcal{T}_{t_s}$ and are candidates of the next to-be-added leaf for the $(t_s + 1)$-th iteration of Algorithm 3. We denote these leaves as $v_k^{(1)}, v_k^{(2)}, \ldots, v_k^{(w)}$. If all these leaves can be added to $\mathcal{T}_{t_s}$ successfully, we will obtain a $\mathcal{T}_{t_{s+1}}$ that is a transcribed tree of $\mathcal{T}_{x_i,s+1}$, which completes the proof. So the remaining question is how to add all of them successfully to $\mathcal{T}_{t_s}$.

For the $(t_s + 1)$-th iteration, we design the LF module in Line 4 to choose $v_k^{(1)}$, based on which the iteration proceeds. If $\mathsf{yca}(v_k^{(1)})$ is a check node, no pivoting will be performed and $v_k^{(1)}$ is added to $\mathcal{T}_{t_s}$ directly without any unexpected side effect. $(w-1)$ leaves $v_k^{(2)}, \ldots, v_k^{(w)}$ remain. For the case in which $\mathsf{yca}(v_k^{(1)})$ is a variable node, a pivoting will be performed and subtrees will be duplicated. Some nodes in the previous remaining list $v_k^{(2)}, \ldots, v_k^{(w)}$ are duplicated as well, and the overall effect may be a list $v_k^{(2)}, \ldots, v_k^{(w)}, v_k^{(w+1)}, \ldots, v_k^{(w+w_d)}$ longer than the original one. We use the last $w_d$ nodes to denote the duplicated leaves resulting from pivoting. This node duplication effect presents the major challenge of adding all $\{v_k^{(w)}\}$ into $\mathcal{T}_{t_s}$. The design goal is to have a LF module that does not result in an ever-increasing list of $\{v_k^{(w)}\}$.

Without loss of generality, we assume that $v_k^{(2)}, \ldots, v_k^{(1+w_d)}$ are the $w_d$ leaves in $\overline{\mathcal{T}}_{t_s}$ that will be duplicated by the left and the right subtrees in Fig. 4(b). After pivoting, the original leaves are duplicated and have identical copies in the left and the right subtrees. We use the first $w_d$ leaves $v_k^{(2)}, \ldots, v_k^{(1+w_d)}$ to represent the corresponding leaves in the right subtree, and use $v_k^{(w+1)}, \ldots, v_k^{(w+w_d)}$ for the corresponding leaves in the left subtree. For the $(t_s+2)$-th iteration, the LF module locates $v_k^{(w+1)}$ as the next to-be-added leaf. Since the root of the left subtree, denoted as $v$, is an active variable node with free observation $y_k$, $\mathsf{yca}(v_k^{(w+1)})$ is simply $v$. By invoking Rule 4, no pivoting is necessary and we can directly include $v_k^{(w+1)}$ by setting its incoming observation $y_k = 1$. By adding the remaining $v_k^{(w+2)}, \ldots, v_k^{(w+w_d)}$ into $\mathcal{T}_{t_s+1}$ sequentially, and noting that only Rule 4 will be invoked, we are able to shortened the list of corresponding leaves to $v_k^{(2)}, \ldots, v_k^{(w)}$. In sum, the end result of the above construction is that regardless of whether $\mathsf{yca}(v_k^{(1)})$ is a variable node or not, this LF module is able to successfully add $v_k^{(1)}$ into $\mathcal{T}_{t_s}$ without introducing any new leaves. Repeating the same procedure for $v_k^{(2)}, \ldots, v_k^{(w)}$, we have constructed a LF module that generates $\mathcal{T}_{t_{s+1}}$, a transcribed tree of $\mathcal{T}_{x_i,s}$. The LF module under this construction is thus balanced growing. The proof is complete. ∎

*Proof of Theorem 4:* With the concept of balanced-growing LF modules, the proof of *Theorem 4* becomes straightforward. We construct a series of finite trees $\{\mathcal{T}_{x_i,0}, \ldots, \mathcal{T}_{x_i,s}, \ldots\}$ by the breadth first search. Its corresponding balanced growing LF module is then one optimal

LF module satisfying *Theorem 4*, the justification of which is as follows.

Let $s_0$ denote the first $s$ such that the breadth first search has visited all nodes of depth of depth $2n$. Since the iterative decoder stops after at most $n$ rounds of message exchanging, the finite tree $\mathcal{T}_{x_i,s_0}$ is equivalent to the infinite tree $\overline{\mathcal{T}}_i$ in terms of function outputs. The balanced growing LF module guarantees that after a finite time $t_0 \stackrel{\Delta}{=} t_{s_0}$, $\mathcal{T}_{t_0}$ is a transcribed tree of $\mathcal{T}_{x_i,s_0}$. Therefore $f_{\mathcal{T}_{t_0}}(\mathbf{y}) = f_{\mathcal{T}_{x_i,s_0}}(\mathbf{y}) = f_{\overline{\mathcal{T}}_{x_i}}(\mathbf{y}) = f_i(\mathbf{y})$ for all $\mathbf{y}$, which implies that $\mathbf{X}_{t_0}$ is the collection of all stopping sets involving $x_i$. By *Theorem 3*, $\forall t \geq t_0$, $\mathbf{X}_t$ is the collection of all stopping sets containing $x_i$, and the proof is complete. ∎

## REFERENCES

[1] J. Pearl, *Probabilistic Reasoning in Intelligent Systems: Network of Plausible Inference*. San Mateo, CA: Morgan Kaufmann, 1988.
[2] R. G. Gallager, *Low-Density Parity-Check Codes*, ser. Research Monograph Series. Cambridge, MA: MIT Press, 1963, no. 21.
[3] D. J. C. MacKay, "Good error-correcting codes based on very sparse matrices," *IEEE Trans. Inform. Theory*, vol. 45, no. 2, pp. 399–431, Mar. 1999.
[4] C. Di, D. Proietti, E. Telatar, T. J. Richardson, and R. L. Urbanke, "Finite-length analysis of low-density parity-check codes on the binary erasure channel," *IEEE Trans. Inform. Theory*, vol. 48, no. 6, pp. 1570–1579, June 2002.
[5] J. Zhang and A. Orlitsky, "Finite length analysis of large- and irregular-left degree ldpc codes over erasure channels," in *Proc. IEEE Int'l. Symp. Inform. Theory*. Lausanne, Switzerland, 2002, p. 3.
[6] T. Richardson, "Error floors of LDPC codes," in *Proc. 41st Annual Allerton Conf. on Comm., Contr., and Computing*. Monticello, IL, 2003.
[7] D. J. C. MacKay and M. S. Postol, "Weakness of Margulis and Ramanujan-Margulis low-density parity check codes," *Electronic Notes in Theoretical Computer Science*, vol. 74, 2003.
[8] P. Vontobel and R. Koetter, "Graph-cover decoding and finite-length analysis of message-passing iterative decoding of LDPC codes." *IEEE Trans. Inform. Theory*, submission preprint - arXiv:cs.IT/0512078.
[9] M. G. Stepanov and M. Chertkov, "Instanton analysis of low-density parity-check codes in the error-floor regime," in *Proc. IEEE Int'l Symp. Inform. Theory*. Seattle, WA, July 2006, pp. 552–556.
[10] J. S. Yedidia, E. B. Sudderth, and J.-P. Bouchaud, "Projection algebra analysis of error correcting codes," Mitsubishi Electric Research Laboratories, Technical Report TR2001-35, 2001.
[11] T. J. Richardson and R. L. Urbanke, "The capacity of low-density parity-check codes," *IEEE Trans. Inform. Theory*, vol. 47, no. 2, pp. 599–618, Feb. 2001.
[12] C. C. Wang, S. R. Kulkarni, and H. V. Poor, "Density evolution for asymmetric memoryless channels," *IEEE Trans. Inform. Theory*, vol. 51, no. 12, pp. 4216–4236, Dec. 2005.
[13] A. Amraoui, R. Urbanke, A. Montanari, and T. Richardson, "Further results on finite-length scaling for iteratively decoded LDPC ensembles," in *Proc. IEEE Int'l. Symp. Inform. Theory*. Chicago, IL, 2004.
[14] C. Cole, S. Wilson, E. Hall, and T. Giallorenzi, "Analysis and design of moderate length regular LDPC codes with low error floors," in *Proc. 40th Conf. Inform. Sciences and Systems*. Princeton, NJ, March 2006.
[15] R. Holzlöhner, A. Mahadevan, C. R. Menyuk, J. M. Morris, and J. Zweck, "Evaluation of the very low BER of FEC codes using dual adaptive importance sampling," *IEEE Commun. Letters*, vol. 9, no. 2, pp. 163–165, Feb. 2005.
[16] M. G. Stepanov, V. Chernyak, M. Chertkov, and B. Vasic, "Diagnosis of weaknesses in modern error correction codes: a physics approach," *Phys. Rev. Lett.*, no. 95, 2005, extended version with supplemental materials - arXiv.org:cond-mat/0506037.
[17] T. Richardson, M. A. Shokrollahi, and R. L. Urbanke, "Finite-length analysis of various low-density parity-check ensembles for the binary erasure channel," in *Proc. IEEE Int'l. Symp. Inform. Theory*. Lausanne, Switzerland, 2002, p. 1.
[18] C.-C. Wang, "Code annealing and the suppressing effect of the cyclically lifted LDPC code ensemble," in *2006 IEEE Information Theory Workshop*. Chengdu, China, October 2006.
[19] A. Orlitsky, K. Viswanathan, and J. Zhang, "Stopping set distribution of LDPC code ensembles," *IEEE Trans. Inform. Theory*, vol. 51, no. 3, pp. 929–953, March 2005.
[20] S. Litsyn and V. Shevelev, "On ensembles of low-density parity-check codes: Asymptotic distance distributions," *IEEE Trans. Inform. Theory*, vol. 48, no. 4, pp. 887–908, Apr. 2002.
[21] ——, "Distance distributions in ensembles of irregular low-density parity-check codes," *IEEE Trans. Inform. Theory*, vol. 49, no. 12, pp. 3140–3159, Dec. 2003.
[22] M. Schwartz and A. Vardy, "On the stopping distance and the stopping redundancy of codes," *IEEE Trans. Inform. Theory*, vol. 52, no. 3, pp. 922–932, March 2006.
[23] K. M. Krishnan and P. Shankar, "On the complexity of finding stopping distance in Tanner graphs," preprint.
[24] E. Berlekamp, R. McEliece, and H. van Tilborg, "On the inherent intractability of certain coding problems," *IEEE Trans. Inform. Theory*, vol. 24, no. 3, pp. 384–386, March 1978.
[25] I. Dumer, D. Micciancio, and M. Sudan, "Hardness of approximating the minimum distance of a linear code," *IEEE Trans. Inform. Theory*, vol. 49, no. 1, pp. 22–37, Jan. 2003.
[26] A. Vardy, "The intractability of computing the minimum distance of a code," *IEEE Trans. Inform. Theory*, vol. 43, no. 11, pp. 1757–1766, Nov. 1997.
[27] K. Abdel-Ghaffar and J. Weber, "Stopping set enumerators of full-rank parity-check matrices of Hamming codes," in *Proc. IEEE Int'l Symp. Inform. Theory*. Seattle, WA, July 2006, pp. 1544–1548.
[28] M. Fossorier, "Quasi-cyclic low-density parity-check codes from circulant permutation matrices," *IEEE Trans. Inform. Theory*, vol. 50, no. 8, pp. 1788–1793, Aug. 2000.
[29] R. Tanner, D. Sridhara, A. Sridharan, T. Fuja, and D. Costello, Jr., "LDPC block and convolutional codes based on circulant matrices," *IEEE Trans. Inform. Theory*, vol. 50, no. 12, pp. 2966–2984, Dec. 2004.
[30] R. Smarandache, A. Pusane, P. Vontobel, and D. Costello, Jr., "Pseudo-codewords in LDPC convolutional codes," in *Proc. IEEE Int'l. Symp. Inform. Theory*. Seattle, WA, July 2006, pp. 1364–1368.
[31] C. Berrou and S. Vaton, "Computing the minimum distance of linear codes by the error impulse method," in *Proc. IEEE Int'l. Symp. Inform. Theory*. Luasanne, Switzerland, July 2002, p. 5.
[32] X. Hu, M. Fossorier, and E. Eleftheriou, "On the computation of the minimum distance of low-density parity-check codes," in *Proc. IEEE Int'l. Conf. Commun.* Paris, France, 2004, pp. 767–771.
[33] Y. Mao and A. H. Banihashemi, "A heuristic search for good low-density parity-check codes at short block lengths," in *Proc. IEEE Int'l. Conf. Commun.* Helsinki, Finland, June 2001.
[34] X. Hu, E. Eleftheriou, and D. M. Arnold, "Regular and irregular progressive edge-growth Tanner graphs," *IEEE Trans. Inform. Theory*, vol. 51, no. 1, pp. 386–398, January 2005.
[35] T. Tian, C. Jones, J. Villasenor, and R. Wesel, "Selective avoidance of cycles in irregular LDPC code construction," *IEEE Trans. Commun.*, vol. 52, no. 8, pp. 1242–1247, August 2004.
[36] A. Ramamoorthy and R. Wesel, "Construction of short block length irregular low-density parity-check codes," in *Proc. ICC*, 2004.
[37] E. Sharon and S. Litsyn, "A method for constructing LDPC codes with low error floor," in *Proc. IEEE Int'l. Symp. Inform. Theory*, July 2006, pp. 2569–2573.
[38] J. Tillich and G. Zemor, "On the minimum distance of structured LDPC codes with two variable nodes of degree 2 per parity-check equation," in *Proc. IEEE Int'l. Symp. Inform. Theory*. Seattle, WA, July 2006, pp. 1549–1553.
[39] P. Vontobel and R. Koetter, "Lower bounds on the minimum pseudo-weight of linear codes," in *Proc. IEEE Int'l. Symp. Inform. Theory*. Chicago, IL, 2004.
[40] X. Wu, X. You, and C. Zhao, "An efficient girth-locating algorithm for quasi-cyclic LDPC codes," in *Proc. IEEE Int'l. Symp. Inform. Theory*. Seattle, WA, 2006, p. 30.
[41] T. Halford, K. Chugg, and A. Grant, "Which codes have 4-cycle-free Tanner graphs," in *Proc. IEEE Int'l. Symp. Inform. Theory*. Seattle, WA, 2006, pp. 871–875.
[42] T. Etzion, A. Trachtenberg, and A. Vardy, "Which codes have cycle-free Tanner graphs," *IEEE Trans. Inform. Theory*, vol. 45, no. 6, pp. 2173–2181, Sept. 1999.
[43] S. Hoory, "The size of bipartite graphs with a given girth," *Journal of Combinatorial Theory. Series B*, vol. 86, no. 2, pp. 215–220, 2002.
[44] M. Twitto, I. Sason, and S. Shamai, "Tightened upper bounds on the ML decoding error probability of binary linear codes," *IEEE Trans. Inform. Theory*, submission preprint.

[45] S. Shamai and I. Sason, "Variations on the Gallager bounds, connections, and applications," *IEEE Trans. Inform. Theory*, vol. 48, no. 12, pp. 3029–3051, Dec. 2002.
[46] J. Rosenthal and P. Vontobel, "Constructions of LDPC codes using Ramanujan graphs and ideas from Margulis," in *Proc. of 38th Allerton Conf. Commun. Control & Computing*. Monticello, IL, 2000, pp. 248–257.
[47] G. Margulis, "Explict constructions of graphs without short cycles," *Combinatorica*, vol. 2, no. 1, pp. 71–78, 1982.
[48] F. R. Kschischang, B. J. Frey, and H.-A. Loeliger, "Factor graphs and the sum-product algorithm," *IEEE Trans. Inform. Theory*, vol. 47, no. 2, pp. 498–519, Feb. 2001.
[49] "P vs. NP," in *Millennium Problems*. The Clay Mathematics Institute of Cambridge, Massachusetts, http://www.claymath.org/millennium/P_vs_NP/.
[50] M. Sipser, *Introduction to the Theory of Computation, Second Edition*. Course Technology, 2005, ISBN: 0-534-95097-3.
[51] J. S. Yedidia and J.-P. Bouchaud, "Renormalization group approach to error-correcting codes," Mitsubishi Electric Research Laboratories, Technical Report TR2001-19, 2001.
[52] D. Divsalar, "Ensemble weight enumerators for protograph LDPC codes," in *Proc. IEEE Int'l Symp. Inform. Theory*. Seattle, WA, July 2006.
[53] http://lthcwww.epfl.ch/research/.
[54] http://magma.maths.usyd.edu.au/.